\documentclass[aps,prc,superscriptaddress,showpacs,amsfonts,twocolumn,floatfix]{revtex4-1}

\usepackage{color} 
\usepackage{amsmath,amssymb} 
\usepackage{graphicx}
\usepackage{lineno}

\definecolor{gray01}{gray}{0.9}
\definecolor{gray02}{gray}{0.8}
\definecolor{gray03}{gray}{0.7}
\definecolor{gray04}{gray}{0.6}
\definecolor{gray05}{gray}{0.5}
\definecolor{gray06}{gray}{0.4}
\definecolor{gray07}{gray}{0.3}
\definecolor{gray08}{gray}{0.2}
\definecolor{gray09}{gray}{0.1}

\newcommand{\bl}[1]{{\color{blue} #1}}
\newcommand{\re}[1]{{\color{red} #1}}

\begin{document}

\title{Measurement of the beam asymmetry~$\Sigma$ and the target asymmetry~$T$ in the 
photoproduction of $\omega$~mesons off the proton using CLAS at Jefferson Laboratory}

\newcommand*{\ANL}{Argonne National Laboratory, Argonne, Illinois 60439}
\newcommand*{\ANLindex}{1}
\affiliation{\ANL}
\newcommand*{\ASU}{Arizona State University, Tempe, Arizona 85287-1504}
\newcommand*{\ASUindex}{2}
\affiliation{\ASU}
\newcommand*{\BONN}{Helmholtz-Institut f\"ur Strahlen- und Kernphysik, Universit\"at Bonn, 53115 Bonn, Germany}
\affiliation{\BONN}
\newcommand*{\CSUDH}{California State University, Dominguez Hills, Carson, CA 90747}
\newcommand*{\CSUDHindex}{3}
\affiliation{\CSUDH}
\newcommand*{\CANISIUS}{Canisius College, Buffalo, NY}
\newcommand*{\CANISIUSindex}{4}
\affiliation{\CANISIUS}
\newcommand*{\CMU}{Carnegie Mellon University, Pittsburgh, Pennsylvania 15213}
\newcommand*{\CMUindex}{5}
\affiliation{\CMU}
\newcommand*{\CUA}{Catholic University of America, Washington, D.C. 20064}
\newcommand*{\CUAindex}{6}
\affiliation{\CUA}
\newcommand*{\SACLAY}{IRFU, CEA, Universit\'e Paris-Saclay, F-91191 Gif-sur-Yvette, France}
\newcommand*{\SACLAYindex}{7}
\affiliation{\SACLAY}
\newcommand*{\CNU}{Christopher Newport University, Newport News, Virginia 23606}
\newcommand*{\CNUindex}{8}
\affiliation{\CNU}
\newcommand*{\UCONN}{University of Connecticut, Storrs, Connecticut 06269}
\newcommand*{\UCONNindex}{9}
\affiliation{\UCONN}
\newcommand*{\FU}{Fairfield University, Fairfield CT 06824}
\newcommand*{\FUindex}{10}
\affiliation{\FU}
\newcommand*{\FERRARAU}{Universit\`a di Ferrara, 44121 Ferrara, Italy}
\newcommand*{\FERRARAUindex}{11}
\affiliation{\FERRARAU}
\newcommand*{\FIU}{Florida International University, Miami, Florida 33199}
\newcommand*{\FIUindex}{12}
\affiliation{\FIU}
\newcommand*{\FSU}{Florida State University, Tallahassee, Florida 32306}
\newcommand*{\FSUindex}{13}
\affiliation{\FSU}
\newcommand*{\NRC}{NRC ``Kurchatov Institute", PNPI, 188300, Gatchina, Russia}
\affiliation{\NRC}
\newcommand*{\GENOVAU}{Universit\`a di Genova, Dipartimento di Fisica, 16146 Genova, Italy}
\affiliation{\GENOVAU}
\newcommand*{\GWUI}{The George Washington University, Washington, DC 20052}
\newcommand*{\GWUIindex}{14}
\affiliation{\GWUI}
\newcommand*{\ISU}{Idaho State University, Pocatello, Idaho 83209}
\newcommand*{\ISUindex}{15}
\affiliation{\ISU}
\newcommand*{\INFNFE}{INFN, Sezione di Ferrara, 44100 Ferrara, Italy}
\newcommand*{\INFNFEindex}{16}
\affiliation{\INFNFE}
\newcommand*{\INFNFR}{INFN, Laboratori Nazionali di Frascati, 00044 Frascati, Italy}
\newcommand*{\INFNFRindex}{17}
\affiliation{\INFNFR}
\newcommand*{\INFNGE}{INFN, Sezione di Genova, 16146 Genova, Italy}
\newcommand*{\INFNGEindex}{18}
\affiliation{\INFNGE}
\newcommand*{\INFNRO}{INFN, Sezione di Roma Tor Vergata, 00133 Rome, Italy}
\newcommand*{\INFNROindex}{19}
\affiliation{\INFNRO}
\newcommand*{\INFNTUR}{INFN, Sezione di Torino, 10125 Torino, Italy}
\newcommand*{\INFNTURindex}{20}
\affiliation{\INFNTUR}
\newcommand*{\ORSAY}{Institut de Physique Nucl\'eaire, CNRS/IN2P3 and Universit\'e Paris Sud, Orsay, France}
\newcommand*{\ORSAYindex}{21}
\affiliation{\ORSAY}
\newcommand*{\ITEP}{Institute of Theoretical and Experimental Physics, Moscow, 117259, Russia}
\newcommand*{\ITEPindex}{22}
\affiliation{\ITEP}
\newcommand*{\JMU}{James Madison University, Harrisonburg, Virginia 22807}
\newcommand*{\JMUindex}{23}
\affiliation{\JMU}
\newcommand*{\JINR}{Joint Institute for Nuclear Research, 141980 Dubna, Russia}
\affiliation{\JINR}
\newcommand*{\KNU}{Kyungpook National University, Daegu 41566, Republic of Korea}
\newcommand*{\KNUindex}{24}
\affiliation{\KNU}
\newcommand*{\MISS}{Mississippi State University, Mississippi State, MS 39762-5167}
\newcommand*{\MISSindex}{25}
\affiliation{\MISS}
\newcommand*{\UNH}{University of New Hampshire, Durham, New Hampshire 03824-3568}
\newcommand*{\UNHindex}{26}
\affiliation{\UNH}
\newcommand*{\NSU}{Norfolk State University, Norfolk, Virginia 23504}
\newcommand*{\NSUindex}{27}
\affiliation{\NSU}
\newcommand*{\OHIOU}{Ohio University, Athens, Ohio  45701}
\newcommand*{\OHIOUindex}{28}
\affiliation{\OHIOU}
\newcommand*{\ODU}{Old Dominion University, Norfolk, Virginia 23529}
\newcommand*{\ODUindex}{29}
\affiliation{\ODU}
\newcommand*{\URICH}{University of Richmond, Richmond, Virginia 23173}
\newcommand*{\URICHindex}{30}
\affiliation{\URICH}
\newcommand*{\ROMAII}{Universit\`a di Roma Tor Vergata, 00133 Rome Italy}
\newcommand*{\ROMAIIindex}{31}
\affiliation{\ROMAII}
\newcommand*{\MSU}{Skobeltsyn Institute of Nuclear Physics, Lomonosov Moscow State University, 119234 Moscow, Russia}
\newcommand*{\MSUindex}{32}
\affiliation{\MSU}
\newcommand*{\SCAROLINA}{University of South Carolina, Columbia, South Carolina 29208}
\newcommand*{\SCAROLINAindex}{33}
\affiliation{\SCAROLINA}
\newcommand*{\TEMPLE}{Temple University,  Philadelphia, PA 19122 }
\newcommand*{\TEMPLEindex}{34}
\affiliation{\TEMPLE}
\newcommand*{\JLAB}{Thomas Jefferson National Accelerator Facility, Newport News, Virginia 23606}
\newcommand*{\JLABindex}{35}
\affiliation{\JLAB}
\newcommand*{\UTFSM}{Universidad T\'{e}cnica Federico Santa Mar\'{i}a, Casilla 110-V Valpara\'{i}so, Chile}
\newcommand*{\UTFSMindex}{36}
\affiliation{\UTFSM}
\newcommand*{\EDINBURGH}{Edinburgh University, Edinburgh EH9 3JZ, United Kingdom}
\newcommand*{\EDINBURGHindex}{37}
\affiliation{\EDINBURGH}
\newcommand*{\GLASGOW}{University of Glasgow, Glasgow G12 8QQ, United Kingdom}
\newcommand*{\GLASGOWindex}{38}
\affiliation{\GLASGOW}
\newcommand*{\VT}{Virginia Tech, Blacksburg, Virginia 24061-0435}
\newcommand*{\VTindex}{39}
\affiliation{\VT}
\newcommand*{\VIRGINIA}{University of Virginia, Charlottesville, Virginia 22901}
\newcommand*{\VIRGINIAindex}{40}
\affiliation{\VIRGINIA}
\newcommand*{\WM}{College of William and Mary, Williamsburg, Virginia 23187-8795}
\newcommand*{\WMindex}{41}
\affiliation{\WM}
\newcommand*{\YEREVAN}{Yerevan Physics Institute, 375036 Yerevan, Armenia}
\newcommand*{\YEREVANindex}{42}
\affiliation{\YEREVAN}
 
\newcommand*{\UM}{University of Michigan, Ann Arbor, MI 48109}
\newcommand*{\KAERI}{Korea Atomic Energy Research Institute, Gyeongju-si, 38180, South Korea}
\newcommand*{\NOWINFNGE}{INFN, Sezione di Genova, 16146 Genova, Italy}
\newcommand*{\NOWUK}{University of Kentucky, Lexington, KY 40506}
\newcommand*{\NOWISU}{Idaho State University, Pocatello, Idaho 83209}
\newcommand*{\NOWSCAROLINA}{University of South Carolina, Columbia, South Carolina 29208}
\newcommand*{\SAUDI}{Imam Abdulrahman Bin Faisal University, Industrial Jubail 31961, Saudi Arabia} 


\author{P.~Roy} \altaffiliation[Present address: ]{\UM}\affiliation{\FSU}
\author{Z.~Akbar} \affiliation{\FSU}
\author{S.~Park} \altaffiliation[Present address: ]{\KAERI}\affiliation{\FSU}
\author{V.~Crede} \altaffiliation[Corresponding author: crede@fsu.edu]{}\affiliation{\FSU}
\author{A.~V.~Anisovich} \affiliation{\BONN} \affiliation{\NRC}
\author{I.~Denisenko} \affiliation{\BONN} \affiliation{\JINR}
\author{E.~Klempt} \affiliation{\BONN} \affiliation{\JLAB}
\author{V.~A.~Nikonov} \affiliation{\BONN} \affiliation{\NRC}
\author{A.~V.~Sarantsev} \affiliation{\BONN} \affiliation{\NRC}

\author {K.~P.~Adhikari} 
\affiliation{\MISS}
\affiliation{\ODU}
% ...


\collaboration{The CLAS Collaboration}
\noaffiliation
\date{Received: \today / Revised version:}

\begin{abstract}
The photoproduction of $\omega$~mesons off the proton has been studied in the reaction $\gamma p\to
p\,\omega$ using the CEBAF Large Acceptance Spectrometer (CLAS) and the frozen-spin target (FROST) in 
Hall~B at the Thomas Jefferson National Accelerator Facility. For the first time, the target asymmetry, 
$T$, has been measured in photoproduction from the decay $\omega\to\pi^+\pi^-\pi^0$, using a 
transversely-polarized target with energies ranging from just above the reaction threshold up to 2.8~GeV. 
Significant non-zero values are observed for these asymmetries, reaching about 30-40\,\% in the 
third-resonance region. New measurements for the photon-beam asymmetry, $\Sigma$, are also presented, 
which agree well with previous CLAS results and extend the world database up to 2.1~GeV. These data 
and additional $\omega$~photoproduction observables from CLAS were included in a partial-wave analysis 
within the Bonn-Gatchina framework. Significant contributions from $s$-channel resonance production 
were found in addition to $t$-channel exchange processes.
\end{abstract}

\maketitle

\section{Introduction}
The internal structure of the nucleon gives rise to an excitation spectrum, which is still poorly 
understood within quantum chromodynamics (QCD). Attempts at understanding the spectrum in 
terms of the basic QCD constituents in lattice-QCD have made significant progress in recent 
years~\cite{Edwards:2011jj}. However, quark models based on effective quark degrees of freedom 
still provide important guidance in our searches for baryon resonances. Known as the so-called 
{\it missing baryon resonances}, many more states have been predicted by phenomenological models 
such as the Constituent Quark Models (CQMs) (see e.g. Refs.~\cite{Capstick:2000qj,Loring:2001ky}) and 
approaches based on a chiral Langrangian~\cite{Sarkar:2009kx} than have been observed experimentally. 
The situation is particularly puzzling in the center-of-mass region above 1.7~GeV and the recent 
lattice-QCD calculations are even consistent with the level counting based on $SU(6)\otimes O(3)$ 
symmetry~\cite{Edwards:2011jj}. Much of our knowledge on nucleon resonances was extracted in 
pion-nucleon scattering experiments~\cite{Olive:2016xmw}, but CQMs have suggested that many 
higher-mass states could decouple from the $\pi N$~channel. For this reason, photoproduction has 
long been considered an important approach in studying the systematics of the spectrum. A summary 
of the progress toward understanding the baryon spectrum is given in Refs.~\cite{Klempt:2009pi,Crede:2013sze}.

It is essential to study nucleon resonances in all their possible decay modes to firmly establish their 
existence and to extract their properties. The production of vector mesons is particularly interesting 
since these mesons ($\rho,\,\omega,\,\phi$) carry the same quantum numbers, $J^{PC} = 1^{--}$, as
the photon and therefore, they are expected to play an important role in photoproduction. The Review 
of Particle Physics~\cite{Olive:2016xmw} clearly shows that the vector-meson decay modes have remained 
under-explored in recent years. However, many hitherto unobserved higher-mass $N^\ast$~resonances 
might strongly couple to these decay modes. The study of $\omega$-meson photoproduction is especially
interesting. The reaction has an additional advantage over $I=1$ vector-meson production since it serves 
as an isospin filter. The $\omega$~meson is an isoscalar particle and therefore, the reaction is sensitive 
only to $I=1/2$ (nucleon) resonances. This reduces the complexity of the contributing intermediate states 
and facilitates the search for new resonances. Moreover, the reaction threshold at $E_\gamma = 1109.1$~MeV 
lies in the third-resonance region around $W\approx 1700$~MeV and thus, provides access to higher-mass 
resonances. 

In photoproduction, the cross section for $\omega$~production is represented by $3\times 2\times 
2\times 2 = 24$ complex numbers, representing the three spin states of the $\omega$, the two spin
states of the initial {\it real} photon, as well as the two spin states of the initial and the recoiling 
proton, respectively. By virtue of parity invariance, 12~relations among these amplitudes exist and 
consequently, only 12~independent complex helicity amplitudes or 24~real numbers remain at each energy 
and angle. In the ideal case of no experimental uncertainties, this will require 23~independent 
measurements (allowing for an overall arbitrary phase) at each energy and angle for a complete description. 
Identifying a set of 23~carefully chosen observables for vector mesons and measuring all of them in order 
to achieve a ``complete experiment''~\cite{Pichowsky:1994gh} remains a challenging task. However, it is 
possible to extract useful dynamical information from the experimentally-accessible polarization 
observables. These observables impose constraints on phenomenological models, thereby aiding 
in reducing the ambiguity in the extraction of the resonance contributions to this reaction. 

The present database of $\omega$~photoproduction observables includes cross-section measurements
from various collaborations~\cite{Barth:2003kv,Williams:2009ab,Wilson:2015uoa,Strakovsky:2014wja}, 
spin-density matrix elements (SDMEs)~\cite{Williams:2009ab,Wilson:2015uoa}, the beam asymmetry 
$\Sigma$~\cite{Ajaka:2006bn,Klein:2008aa,Vegna:2013ccx,Collins:2017vev}, and the double-polarization 
observables $E$~\cite{Eberhardt:2015lwa, Akbar:2017uuk} (helicity asymmetry) 
and $G$~\cite{Eberhardt:2015lwa} (correlation between linear-photon and longitudinal-target 
polarization). The importance of polarization observables for our understanding 
of this reaction has frequently been discussed in the literature,
e.g. Refs.~\cite{Klempt:2009pi,Crede:2013sze}. 

Since the $\omega$~meson has the same quantum numbers as the incoming photon, a dominant $t$-channel 
background contribution is expected. The inclusion of polarized SDMEs and the polarization observables 
$\Sigma$, $E$ and $G$ from the CBELSA/TAPS Collaboration played a crucial role in a recent BnGa 
partial-wave analysis~\cite{Denisenko:2016ugz} toward understanding the nature of the $t$-channel 
amplitudes and disentangling them from the $s$-channel resonance contributions. For example, a data 
description with only $t$-channel amplitudes predicted the beam asymmetry to be close to zero, whereas 
experimentally this asymmetry was observed to be significantly bigger and to exceed values of 0.5 across 
the entire incident-photon energy range below 2~GeV. Linear beam polarization allowed the study of the 
production process in more detail and helped separate natural and unnatural parity-exchange conributions 
(e.g.~pomeron and $\pi$~exchange, respectively). A summary of all published results in 
$\omega$~photoproduction can be found in our preceding paper~\cite{Akbar:2017uuk}.
  

In this paper, first-time measurements are presented for the target asymmetry, $T$, as well as results 
for the beam-asymmetry, $\Sigma$, in the photoproduction reaction:
\begin{equation}
\gamma\,p\,\to\,p\,\omega,\quad{\rm where}\quad\,\omega\,\to\,\pi^+\,\pi^-\,\pi^0
\label{equ:omega_reaction}
\end{equation} 
from the CLAS-FROST experiment. These new measurements cover a broad range in photon energies, 
$E_{\gamma} \in [1.1, 2.1]$~GeV for $\Sigma$ and $E_{\gamma} \in [1.2, 2.8]$~GeV for~$T$. The presented 
results on $\Sigma$ allow a comparison with previously published results and serve as a validation for 
the new measurements of the target asymmetry. Moreover, these $\Sigma$~results also represent first-time 
measurements for the energy range $E_{\gamma} \in [1.9, 2.1]$~GeV.

This paper has the following structure. Section~\ref{Section:ExperimentalSetup} describes the CLAS 
(FROST) experimental setup. The data reconstruction and event selection are discussed in 
Section~\ref{Section:Selection} and the technique for extracting the polarization observables is 
described in Section~\ref{Section:Observables}. Experimental results and a discussion of observed 
resonance contributions are presented in Sections~\ref{Section:Results} and \ref{Section:PWA}, 
respectively. The paper ends with a brief summary and an outlook.

\section{The FROST Experimental Setup\label{Section:ExperimentalSetup}}
\begin{figure}[!t]
  \includegraphics[width=0.41\textwidth,height=0.25\textheight]{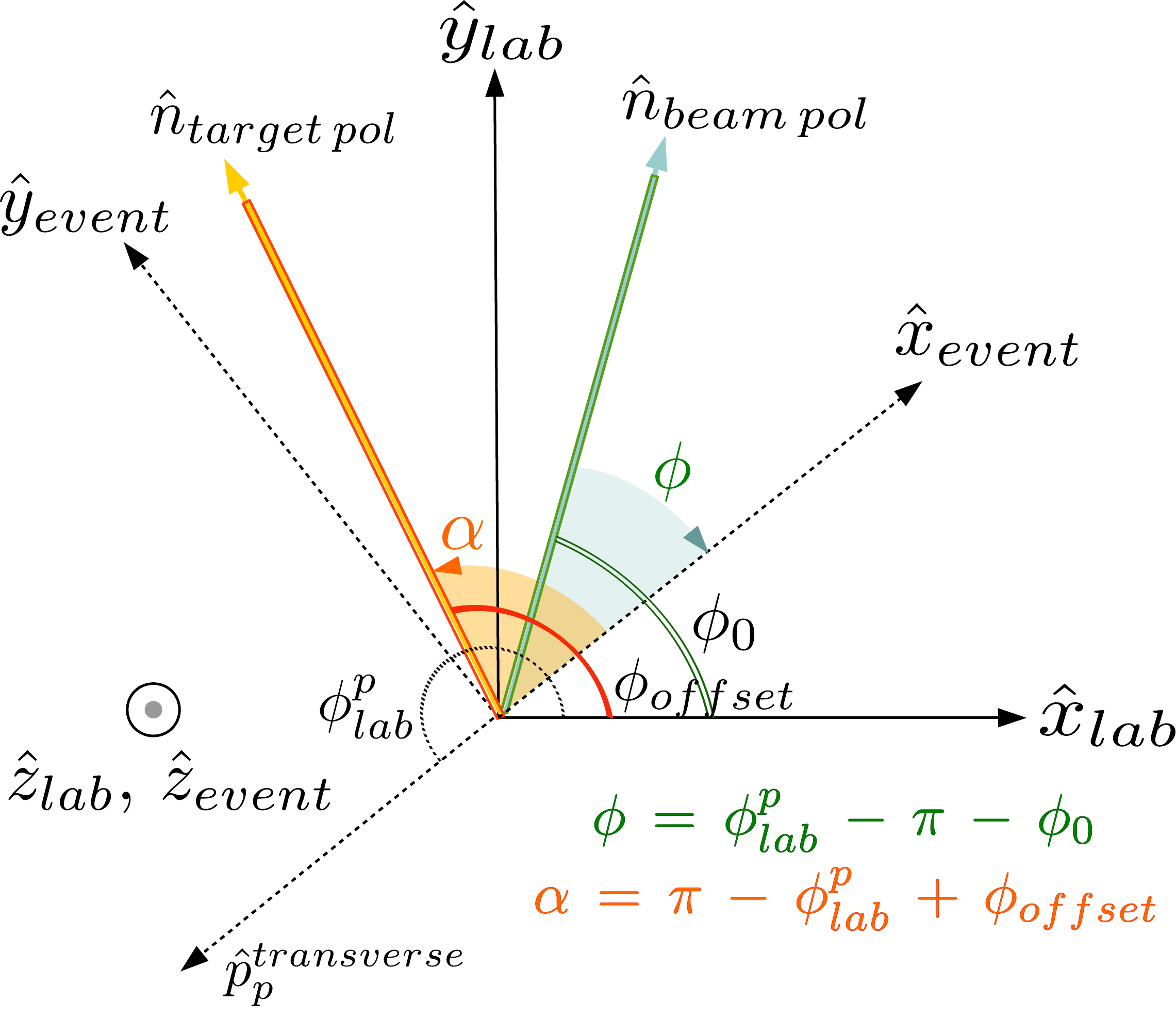}
  \caption{The polarization directions of the linearly-polarized photon beam and the 
     transversely-polarized butanol target in the laboratory and event frames. See text for the definition 
     of the axes. The beam polarization (shown as a green arrow) was inclined at an angle $\phi_0=0^\circ$ 
     with respect to the $x$-axis in the laboratory frame ($\hat{x}_{\rm \,lab}$) for the parallel setting 
     and at $\phi_0=90^{\circ}$ for the perpendicular setting. The target polarization (shown as an orange
     arrow) was inclined at an angle $\phi_{\rm \,offset}$. 
     The picture also shows the azimuthal angle $\phi$ ($\alpha$) of the beam (target) polarization in 
     the event frame and its relation with the azimuthal angle $\phi^{\,p}_{\rm \,lab}$ of the recoiling proton
     in the laboratory frame. More details on how these angles were used in the analysis are discussed in 
     Section~\ref{Section:Observables}.}
     \label{fig:axes_and_angles_def}
\end{figure}

The FROzen Spin Target (FROST)~\cite{Keith:2012ad} experiment was conducted at the Thomas Jefferson 
National Accelerator Facility (Jefferson Lab) in Newport News, Virginia, using the CEBAF Large Acceptance 
Spectrometer (CLAS)~\cite{Mecking:2003zu} in Hall B at Jefferson Lab. FROST covered a variety of individual 
experiments with all possible combinations of linear and circular beam polarization, as well as longitudinal 
and tranverse target polarization, thus providing access to single- and double-polarization observables in a 
large number of reactions~\cite{Strauch:2015zob,Senderovich:2015lek,Akbar:2017uuk}. For these measurements 
of the $\omega$~beam and target asymmetries, the target was transversely polarized and the beam was linearly
as well as circularly polarized, respectively. Figure~\ref{fig:axes_and_angles_def} shows a schematic that 
illustrates the more complex kinematic situation of linear-beam polarization in combination with 
transverse-target polarization in the two coordinate systems relevant for this analysis: the laboratory 
frame and the event frame. The $z$-axis was chosen to be along the direction of the incoming photon beam 
in both frames. The $y$-axis in the laboratory frame, $\hat{y}_{\rm \,lab}$, was chosen along the vertical 
direction pointing away from the floor, and $\hat{x}_{\rm \,lab}$ was given by 
$\hat{x}_{\rm \,lab} = \hat{y}_{\rm \,lab}\,\times\,\hat{z}_{\rm \,lab}$. The $x$- and $y$-axes in the 
event frame were chosen as follows: $\hat{y}_{\rm \,event}$ was perpendicular to the center-of-mass 
production plane. Mathematically, $\hat{y}_{\rm \,event} = (\hat{p}_{\,p}\,\times\,\hat{z}_{\rm \,event}) /
{\left| \hat{p}_{\,p}\,\times\,\hat{z}_{\rm \,event}\right|}$, where $\hat{p}_{\,p}$ is a unit vector along 
the momentum of the recoiling proton in the center-of-mass frame. Then, $\hat{x}_{\rm \,event}$ was given by
$\hat{x}_{\rm \,event} = \hat{y}_{\rm \,event} \times \hat{z}_{\rm \,event}$.

The beam of linearly-polarized tagged photons was created by employing a coherent bremsstrahlung
technique~\cite{Timm:1969mf,Lohmann:1994vz} whereby unpolarized electrons were scattered from a 
$50~{\rm \mu m}$~thick diamond radiator. The electrons were initially accelerated using the Continuous
Electron Beam Accelerator Facility (CEBAF) at Jefferson Lab with energies reaching up to $5.5$~GeV. After 
passing the radiator, the electrons were deflected into a tagging detector system~\cite{Sober:2000we},
which provided the information to tag the time and the energy of the corresponding bremsstrahlung photons 
with a resolution of $\Delta t\sim 100$~ps and $\Delta E/E\approx 0.1\,\%$, respectively.
The orientation of the linear polarization plane could be set to either parallel (denoted as $\parallel$) 
or to perpendicular (denoted as $\perp$) relative to the floor of the experimental hall by adjusting the 
azimuthal angle of the crystal lattice of the diamond radiator~\cite{Livingston:2008hv}. The corresponding 
azimuthal angle of the beam polarization in the laboratory frame was $\phi_0 = 0^{\circ}$ or $90^{\circ}$, 
respectively (see Figure~\ref{fig:axes_and_angles_def}). The angle between a selected diamond plane and the 
incident electron beam determined the leading-edge of an enhancement in the photon energy spectrum known as 
the ``coherent edge.'' The incident photons reached their maximum polarization within a roughly $200$~MeV 
window below the coherent edge. In this experiment using linear beam polarization, coherent-edge settings 
from $0.9$~GeV to $2.1$~GeV in intervals of $0.2$~GeV were used. The average degree of polarization of the 
linearly-polarized beam was measured via a fit of the photon energy spectrum using a coherent bremsstrahlung 
calculation~\cite{Ken_pol_cal} and was found to vary between 40\,-\,60\,\% depending on energy.

For the measurement of the target asymmetry, a circularly-polarized, tagged, bremsstrahlung photon beam 
was used, which
results from a polarization transfer when the incident electron beam itself is longitudinally polarized. Since
the electron beam helicity state flipped rapidly, integrating over the initial helicity states resulted 
effectively in an unpolarized incident photon beam.

The target nucleons were free protons inside a $5$~cm long frozen-spin butanol (C$_4$H$_9$OH) target
system~\cite{Keith:2012ad}. The target was transversely polarized using a Dynamic Nuclear Polarization 
(DNP) technique~\cite{Abragam:1978} outside the CLAS detector in a $5.0$~T homogeneous magnetic field
and at a temperature of $T=200$\,-\,300~mK. To maintain the transverse polarization of the target inside 
the detector system, the target was cooled down to about $60$~mK and a $0.5$~T holding field was applied
using a dipole magnet. An average transverse polarization of about $81\,\%$ was achieved. The polarization
values were determined from regular NMR measurements taken for both target polarizations: pointing away
from the floor (denoted as `$+$') and pointing towards the floor (denoted as `$-$'). The target polarization
was inclined at an angle $\phi_{\rm \,offset}\,=\,116.1^\circ\,\pm\,0.4^{\circ}$ (referred to as the target 
offset angle) from the $x$-axis in the laboratory frame for the `$+$' setting and at 
$\phi_{\rm \,offset}\,=\,-63.9^\circ\,\pm\,0.4^{\circ}$ for the `$-$' setting, as shown in 
Fig.~\ref{fig:axes_and_angles_def}. These offsets were necessary to
prevent photoproduced $e^+e^-$~background from being directed into the CLAS acceptance region by the
target holding field. 

\begin{figure}[t]
  \includegraphics[height=0.26\textheight,width=0.45\textwidth]{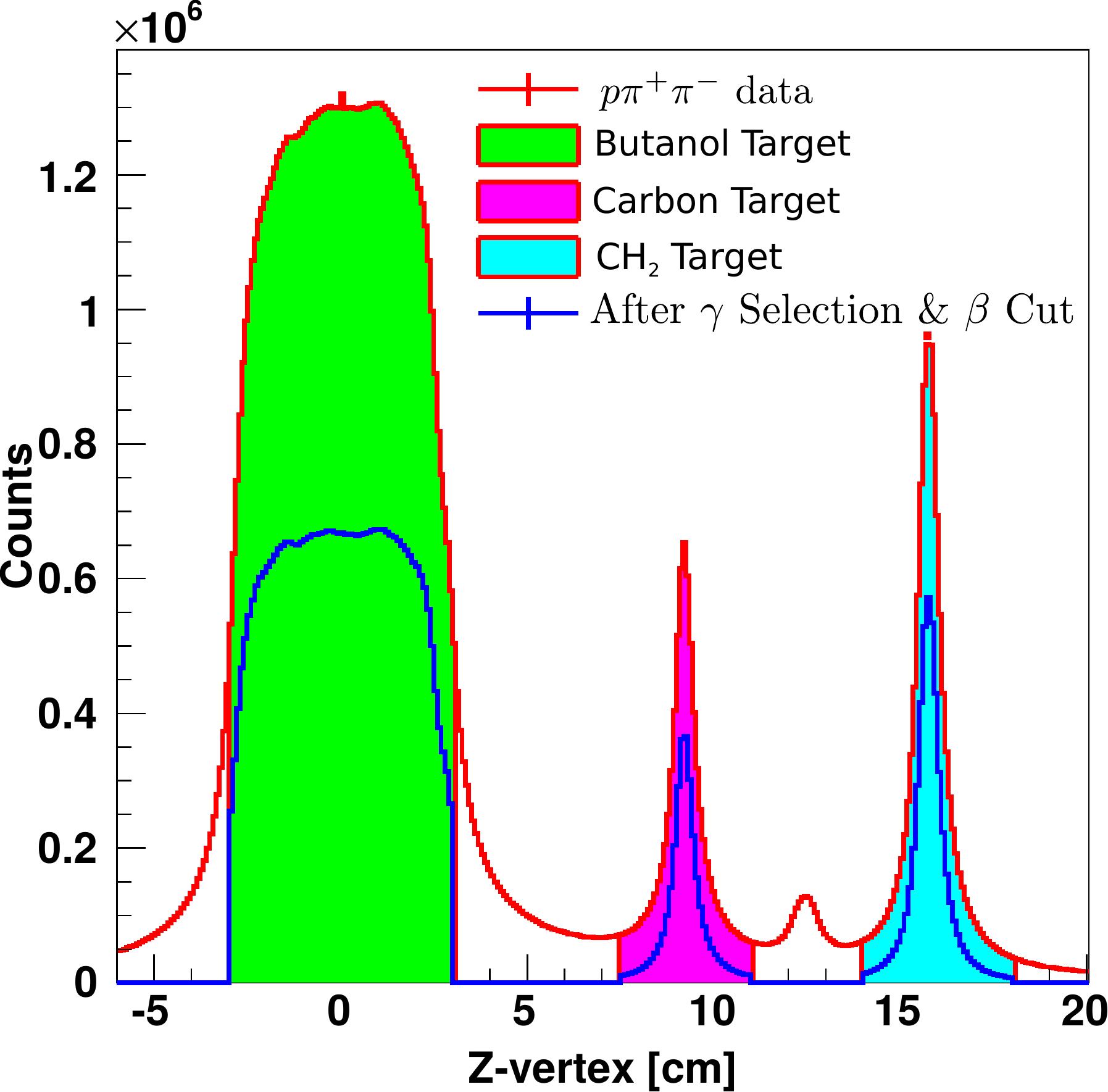}
  \caption{The $z$-vertex distribution (axis along the beam line) of all reconstructed 
      particles. The CLAS center was chosen as the $z=0$ coordinate. The peak on the left shows the $z$ 
      position of the butanol target, the peak situated next to it shows the position of the carbon target 
      and the peak on the right shows the position of the polyethylene (CH$_2$) target. The \re{red} line 
      denotes the data containing all $p\,\pi^+\pi^-$ events. The \bl{blue} line denotes these events after 
      applying photon selection and particle-identification cuts (discussed in 
      Section~\ref{Section:Selection}). The small peak between the carbon and the polyethylene target 
      originated from the end-cap of the heat shield.}
  \label{fig:Vertex_cut}
\end{figure}

In addition to the butanol target, two unpolarized targets were placed in the target cryostat, including 
carbon and polyethylene (CH$_2$) targets for background subtraction and systematic studies. They were 
placed further downstream than the butanol target at approximately $\Delta z = 9$~cm and 16~cm, 
respectively, and were well-separated from each other, as is evident from the $z$-vertex distribution 
shown in Fig.~\ref{fig:Vertex_cut}. The thickness of the additional targets was chosen such that the hadronic 
rate from each was about 10\,\% the rate of butanol.

The charged final-state particles were detected using the CLAS spectrometer~\cite{Mecking:2003zu}, which 
was based on a non-homogeneous toroidal magnetic field, primarily pointing in the $\phi$~direction, with 
a maximum magnitude of $1.8$~T generated by a six-coil torus magnet. The design of the magnet provided 
a field-free region around the polarized target. The CLAS detector system had many components, each with 
a six-fold symmetry about the beam axis, covering a solid angle of about 80\,\% of~$4\pi$. For an event to 
be recorded, the trigger configuration required the detection of at least one charged track.


\section{Event Selection\label{Section:Selection}}
The $\omega$~mesons were reconstructed from their $\pi^+ \pi^- \pi^0$~decay. This decay mode has the 
highest branching fraction of ($89.2\pm 0.7)\,\%$~\cite{Olive:2016xmw}. Events were selected when exactly 
one final-state proton as well as one $\pi^+$ and one $\pi^-$~track were detected. A one-constraint
kinematic fit imposing a missing~$\pi^0$ was used to reconstruct the
four-vector of the neutral pion.

\begin{figure}[b]
  \begin{center}
    \includegraphics[width=0.48\textwidth,height=0.3\textheight]{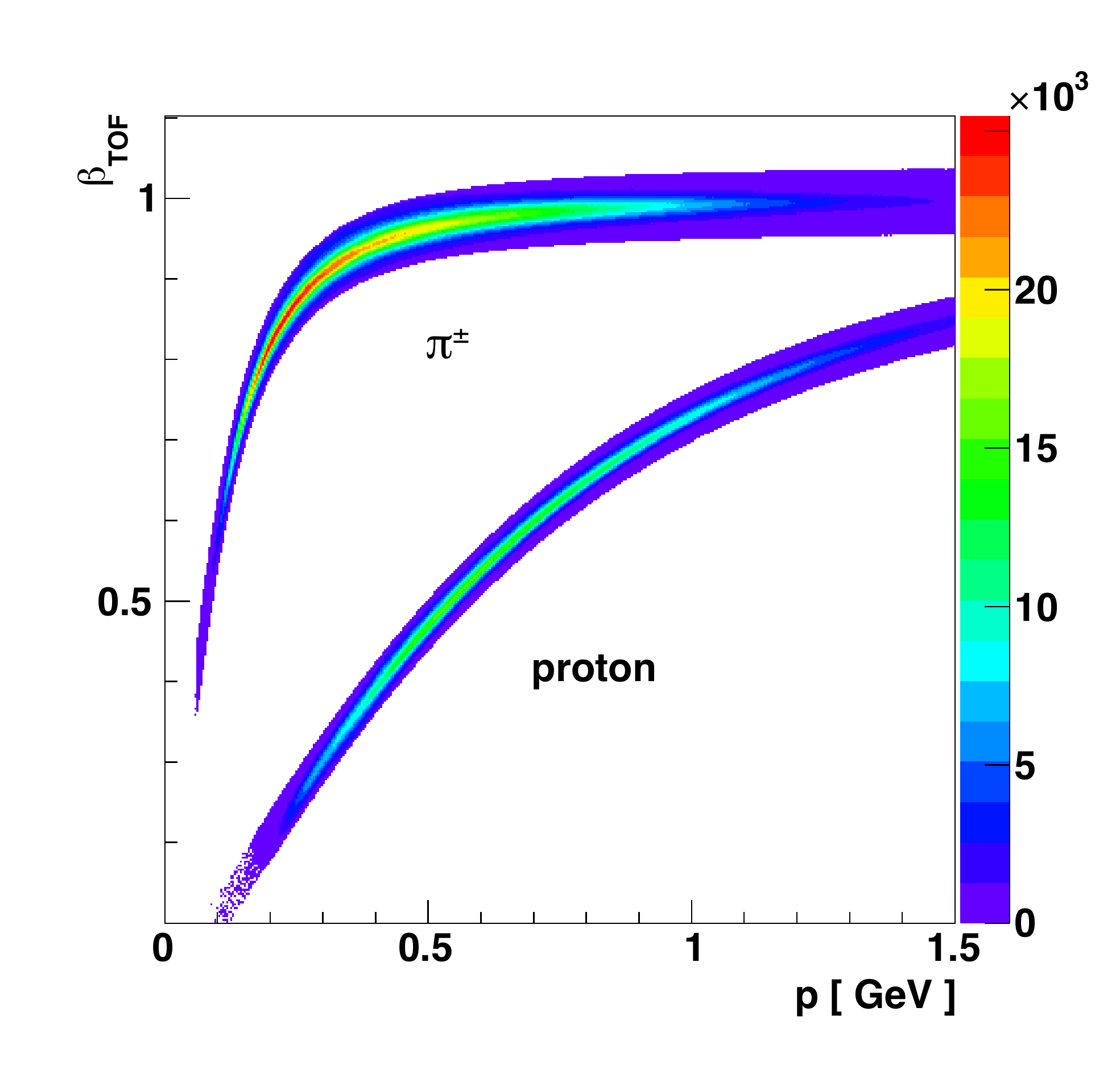}
    \caption{Typical example of a $\beta_{\rm \,TOF}$ versus particle momentum distribution after the 
        $3\sigma$~cuts on $\Delta\beta$.}
    \label{Figure:DeltaBeta}
  \end{center}
\end{figure}

Prior to kinematic fitting, the following cuts and event corrections were applied. Initial photon selection 
cuts were required since the photons arrived at the target in 2~ns bunches. To select the correct photon 
out of several potential candidates, a cut of $\pm 1$~ns on the coincidence time (time difference between 
the event vertex time and the time the photon arrived at the vertex) was applied. This reduced the initial 
situation from approximately five candidate photons per event to only about 8\,-\,10\,\% of all events 
having more than one candidate photon. These events were discarded. To further minimize the ambiguity 
in identifying the correct photon, only those events were considered in which the vertex-timing cut identified 
the same photon for all tracks.

For final-state particle identification, the $\beta$~value of each track was determined from two separate 
sources: (1) $\beta_{\rm \,DC} = p/\sqrt{p^2+m^2}$ was measured using the momentum information from the 
drift chambers and the PDG mass~\cite{Olive:2016xmw} for the particle, and (2) $\beta_{\rm \,TOF}=\frac{v}{c}$ 
used the velocity information from the time-of-flight (TOF) system~\cite{Smith:1999ii, Mecking:2003zu}. 
Events were selected based on $\Delta\beta\,=\,|\beta_{\rm \,DC}\,-\,\beta_{\rm \,TOF}|\,\leq\,3\,\sigma$, 
where $\sigma$ was the width of the Gaussian $\Delta\beta$~distributions, which were centered at zero 
for pions and protons. This led to a significant improvement in the identification of good final-state tracks 
and clear bands for protons and charged pions could be identified in the $\beta_{\,TOF}$ versus momentum
distributions (Fig.~\ref{Figure:DeltaBeta}). In addition, vertex cuts of $x^2 + y^2 < 9$~cm$^2$ and 
$-3.0 < z < 3.0$~cm were applied to select events originating from the butanol target.

\begin{figure}[t]
  \begin{center}
    \includegraphics[width=0.48\textwidth,height=0.29\textheight]{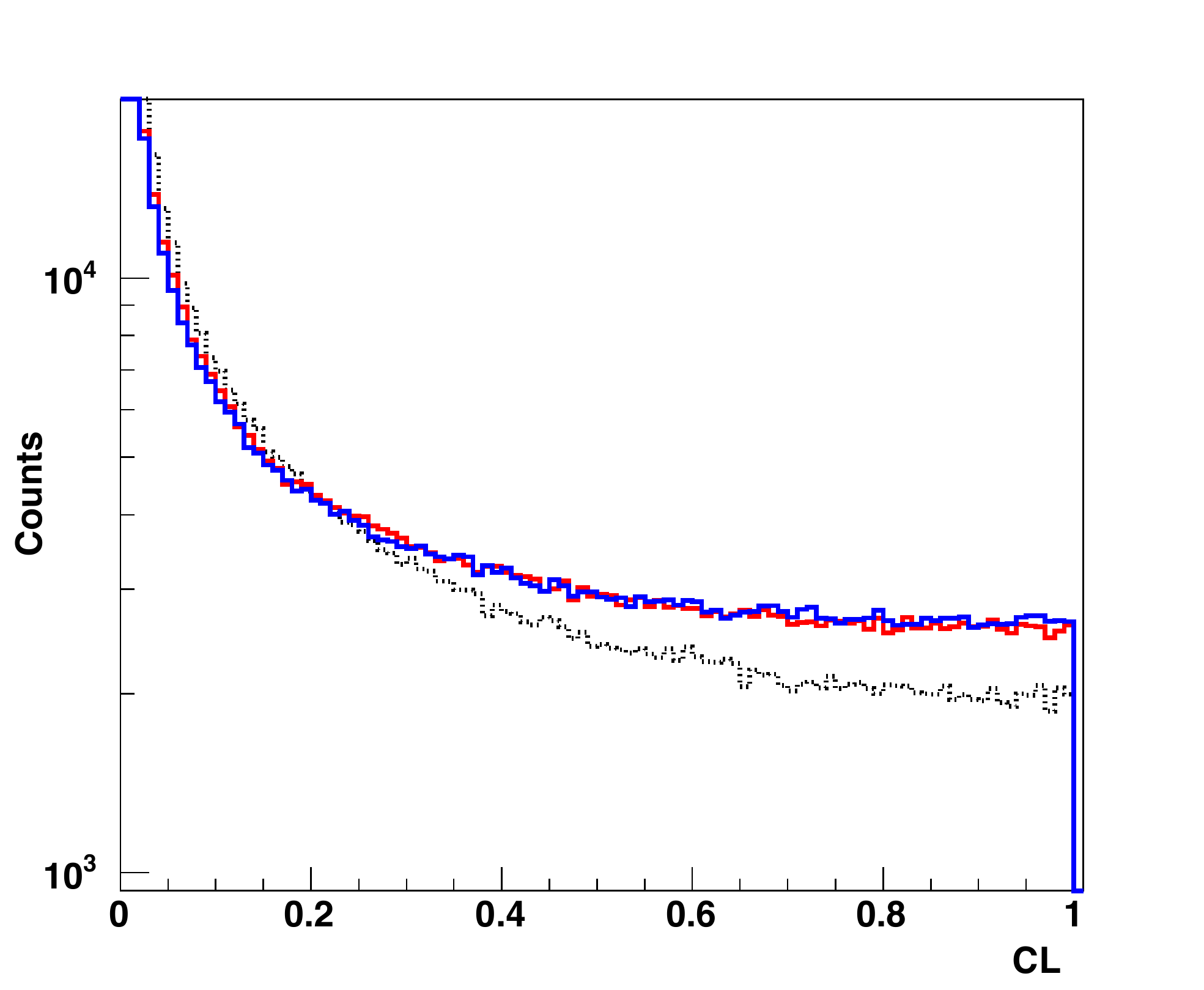}
    \caption{Examples of confidence-level (CL) distributions for the topology $\gamma p\to 
        p\,\pi^+ \pi^-\,(\pi^0)$ from the 1.5-GeV coherent-edge data set for the butanol target. The black 
        dotted line shows the distribution before energy-loss and momentum corrections, the red line before 
        momentum corrections, and the blue line represents the final distribution.}
    \label{fig:CL_Top05}
  \end{center}
\end{figure}

The four-vectors of the selected charged final-state particles were corrected for the energy loss due to 
the interaction with materials while traveling through the CLAS volume. Small momentum corrections of a 
few MeV were also required to correct for factors such as variations in the magnetic field of the torus 
magnet and/or misalignments of the drift chambers. The corrections of the $\pi^+$ and proton four-vectors 
were initially determined such that the mass distributions of $X$ in $\gamma p\to p\,X$ and $\gamma p\to 
p\,\pi^+\,X$ did not have any azimuthal dependence. By using kinematic fitting, these corrections were 
further fine-tuned and momentum-dependent corrections for the $\pi^-$ were also found.

In a final step, a kinematic fit on these corrected four-vectors imposed energy and momentum conservation 
implying a missing $\pi^0$. An example of our confidence-level (CL) distributions is shown in 
Fig.~\ref{fig:CL_Top05}. After applying energy-loss corrections, the slope of the distribution improved 
significantly, approaching the ideal value of zero toward CL = 1. The application of momentum corrections
led to a further improvement in the distribution. However the improvement was small since the momentum 
corrections were much smaller in magnitude than the energy-loss corrections. A very small CL cut of $p > 0.001$ 
was finally applied to simply require fit convergence. This removed most of the $\gamma p\to p\,\pi^+\pi^-$ 
background.

\subsection*{Event-based signal-background separation\label{ssec:Qfac}}
\begin{figure}[!b]
   \includegraphics[width=0.45\textwidth,height=0.26\textheight]{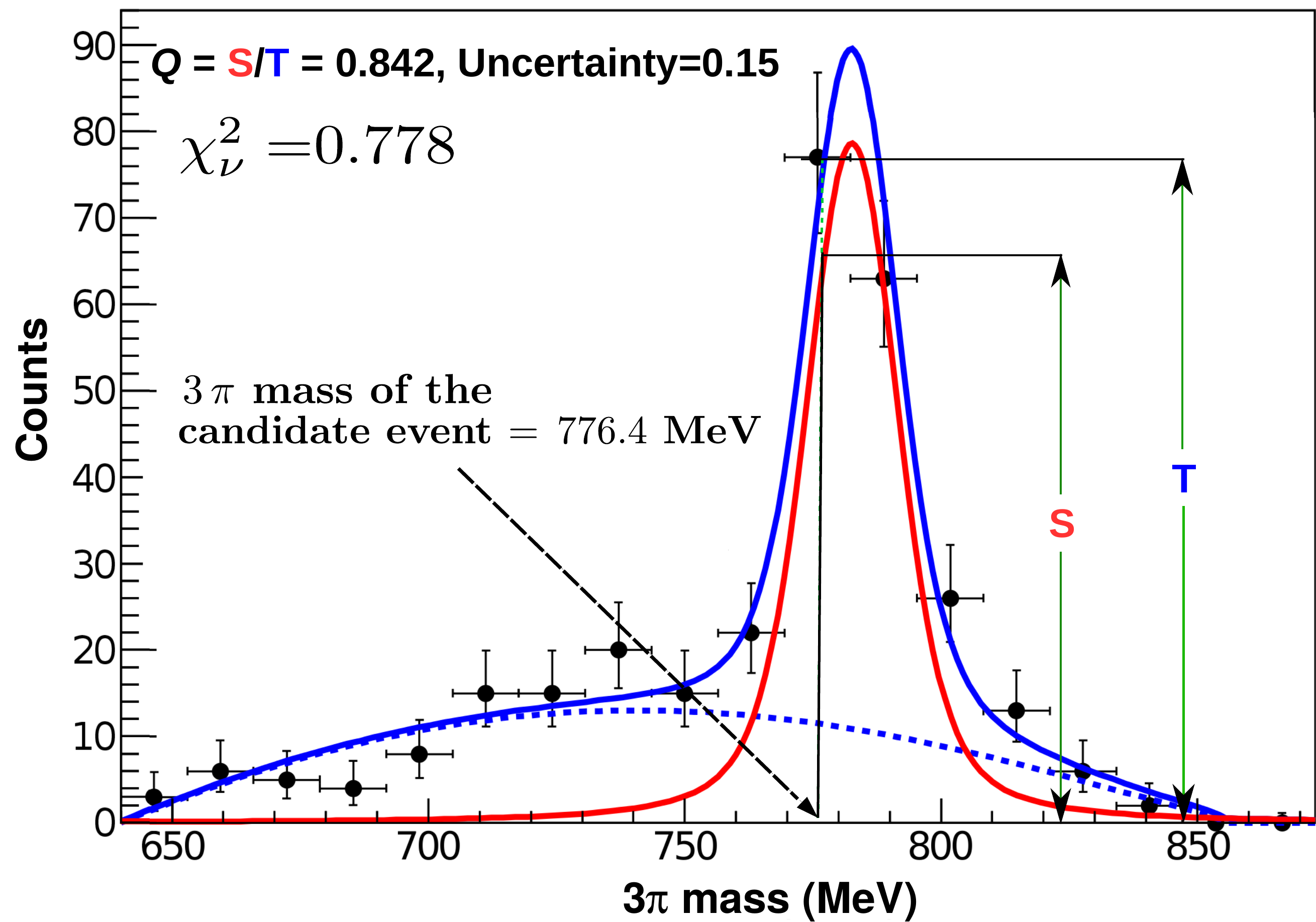}
   \caption{A typical example of a ($\pi^+\pi^-\pi^0$)~mass distribution of the 300~nearest 
       neighbors for an event in the energy bin $E_\gamma \in [1.3,\,1.4]$~GeV. The \bl{blue} solid line represents 
       the total fit, the \re{red} solid line the signal (Voigtian pdf), and the \bl{blue} dotted line the background 
       function (third-order Chebychev pdf). The $Q$~value of the event was given by $Q=S/T$, where $S (T)$ 
       was the height of the signal pdf (total pdf) at the $3\pi$~mass of the candidate event.}
   \label{fig:Qfac_chi2_fit}
\end{figure}
\begin{figure*}[!ht]
   \includegraphics[width=0.49\textwidth,height=0.29\textheight]{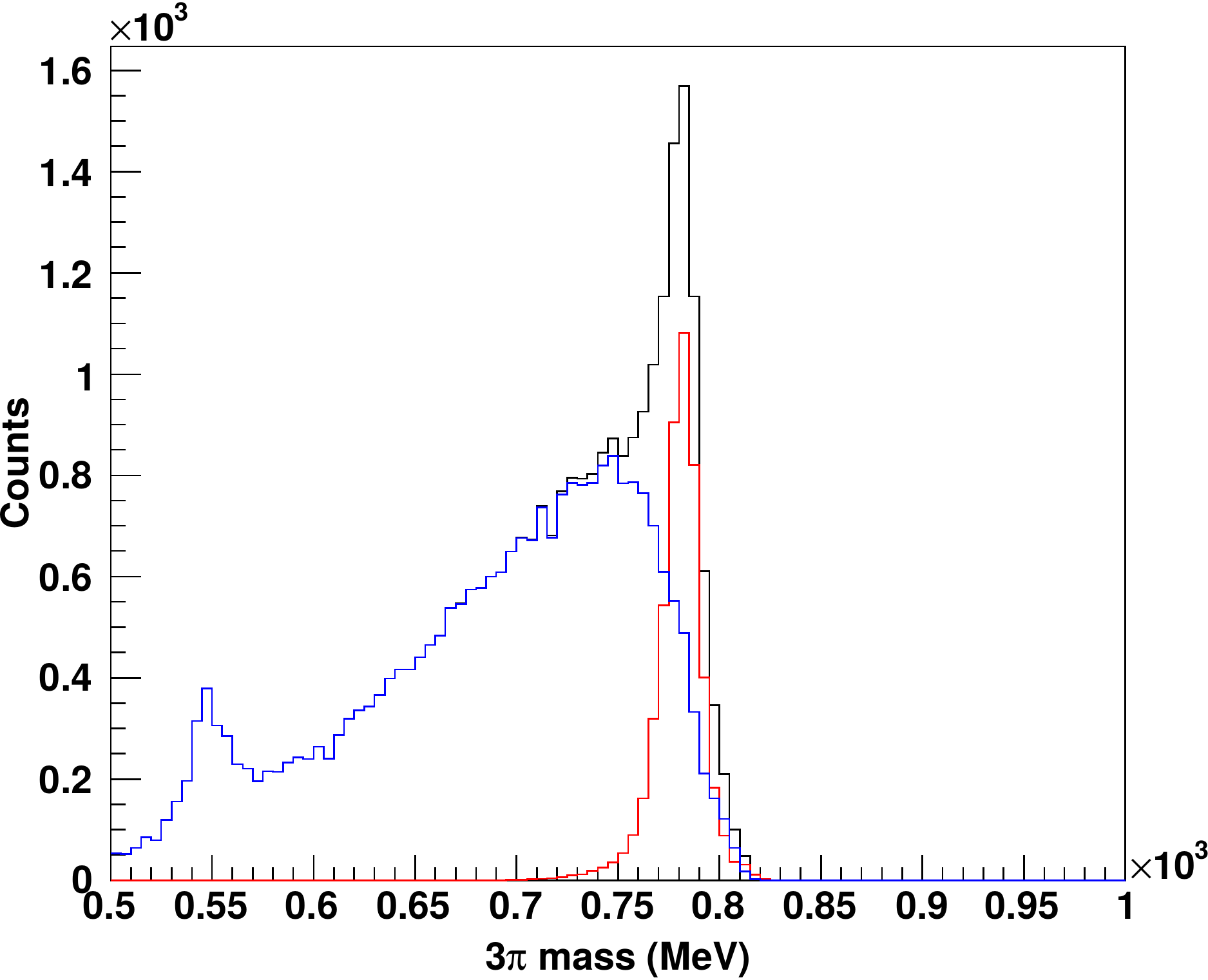}
   \hfill
   \includegraphics[width=0.49\textwidth,height=0.29\textheight]{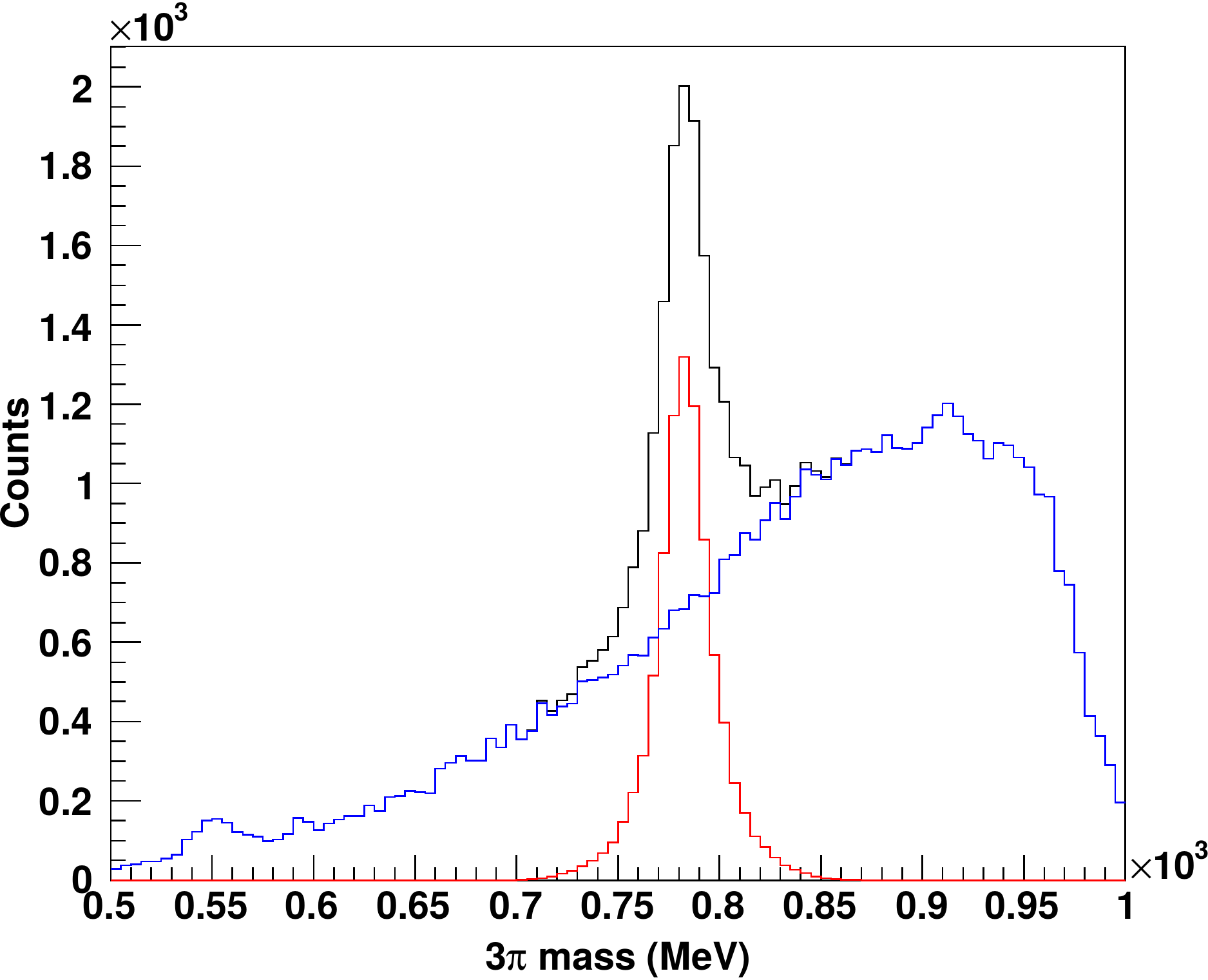}
   \caption{Examples of invariant $\pi^+\pi^-\pi^0$ distributions for 
       $E_\gamma\in [1100,\,1200]$~MeV (left) and $E_\gamma\in [1500,\,1600]$~MeV (right), summed over all 
       angles and all polarization states. The black solid line denotes the full mass distribution, the \re{red} line 
       shows the signal mass distribution obtained after weighting each event with {\it Q} and the \bl{blue} line
       represents the background mass distribution obtained after weighting each event with $(1-Q)$.}
   \label{fig:Qfac_mass_histo}
\end{figure*}
The remaining background, consisting of mostly $p\,\omega$~events originating from bound nucleons of the 
butanol target as well as other non-$p\,\omega$~events resulting in a $p\,\pi^+\pi^-\pi^0$ final state, was 
separated from signal events using a probabilistic event-based method. This multivariate analysis technique 
is described in detail in Ref.~\cite{Williams:2008sh} and its application in previous CLAS analyses on the 
measurement of the $\omega$~photoproduction cross sections and the $\omega$~double-spin asymmetry 
is detailed in Refs.~\cite{Williams:2009ab,Akbar:2017uuk}. The method determines a weight for each event, 
denoted as the event $Q$~value, which denotes the probability for the event being a signal event. These 
$Q$~values were then used as event weights to provide any signal distribution, such as angular or mass 
distributions, and also facilitated the application of event-based likelihood fits (see 
Section~\ref{ssec:Likelihood}). For this method, the data were divided into data subsets based on their photon 
energy (binned in $100$-MeV wide bins) and on their beam and/or target polarization orientations. To determine 
the $Q$~value for each event in any given data subset, the 300 kinematically-nearest neighbors were selected 
using a distance metric in the phase space of all relevant kinematic variables, with the exception of the 
$3\pi$~invariant mass. In this analysis, these variables were: 
\begin{equation*}
{\rm cos}\,\Theta_{\rm \,c.m.}^{\,\omega},\quad {\rm cos}\,\theta_{\rm \,HEL},\quad \phi_{\rm \,HEL},\quad 
  \phi_{\rm \,lab}^{\,\omega}, 
\quad \lambda\,,
\end{equation*}
where ${\rm cos}\,\Theta_{\rm \,c.m.}^{\,\omega}$ denotes the cosine of the polar angle of the $\omega$~meson
in the center-of-mass frame, ${\rm cos}\,\theta_{\rm \,HEL}$ and $\phi_{\rm \,HEL}$ denote the two angles 
of the $\omega$~meson in the helicity frame, $\phi_{\rm \,lab}^{\,\omega}$ is the azimuthal angle of the 
$\omega$~meson in the lab frame, and $\lambda$ is a quantity that is proportional to the $\omega
\to\pi^+\pi^-\pi^0$ decay amplitude~\cite{Williams:2009ab}. It was calculated in terms of the pion momenta 
in the rest frame of the $\omega$:
\begin{equation}
\begin{split}
\lambda\,=\,\frac{|\,\vec{p}_{\,\pi^+}
  \,\times\,\vec{p}_{\,\pi^-}\,|^2}{\lambda_{\rm\,max}},\quad {\rm with~a~maximum~value~of}\\[1.5ex]
\lambda_{\rm \,max} \,=\, T^2\,\bigg(\frac{T^2}{108}\,+\,\frac{mT}{9}\,+\,\frac{m^2}{3} \bigg)
\end{split}
\end{equation}
for a totally symmetric decay, where $T = T_1 + T_2 + T_3$ is the sum of the $\pi^{\pm,\,0}$~kinetic energies 
and $m$ is the $\pi^\pm$~mass. The parameter~$\lambda$ varies between 0 and 1 and shows a linear increase 
as expected for a vector meson.

This method guaranteed the selection of the 300~nearest neighbors in a very small region of the 
multi-dimensional phase space around the candidate event. Therefore, it was assumed that the signal and 
background distributions did not vary rapidly in the selected region and that the $3\pi$~invariant mass 
distribution of these 300~events determined the $Q$~value of the event. Due to the small sample size of the 
selected nearest neighbors, an event-based unbinned maximum likelihood method was implemented to fit 
the mass distributions. The fit function was defined as:
\begin{equation}
  f(x) \,=\, N\, [ f_{s}\,S(x) \,+\, ( 1 \,-\, f_{s} )\, B(x) ]\,,
  \label{equ:total_function}
\end{equation}
where $S(x)$ denoted the signal and $B(x)$ the background probability density function (pdf). $N$ was a 
normalization constant and $f_{s}$ was the signal fraction with a value between 0 and 1. The $Q$~value 
itself was then given by:
\begin{equation}
  Q \,=\, \frac{s(x)}{s(x) \,+\, b(x)}\,,
  \label{equ:Q_factor}
\end{equation}
where $x$ was the $3\pi$ invariant mass of the candidate event, $s(x) = f_{s} \cdot S(x)$ and 
$b(x) = (1-f_{s}) \cdot B(x)$. 

A Voigtian function, which is a convolution of a Gaussian (to describe the resolution) and a Breit-Wigner 
(to describe the natural line shape of the resonance), was chosen to describe the signal pdf. A third-order 
Chebychev polynomial was selected to describe the background pdf. Since the unbinned maximum-likelihood 
fitting technique did not provide any goodness-of-fit measure to check the fit quality, the output of each 
likelihood fit was used to perform a least-squares fit of the $3\pi$-mass distribution of the same 300~events.
The corresponding $\chi^2_\nu$ value provided the goodness of fit. An example of such a least-squares 
fit is shown in Figure~\ref{fig:Qfac_chi2_fit}. The figure also demonstrates how the {\it Q}~value was 
calculated for a candidate event. The choice of a Voigtian for the signal pdf and a third-order Chebychev 
for the background pdf gave the overall best distribution for the reduced-$\chi^2$. For the energy bins 
close to the $\omega$~production threshold, a product of an Argus function and a second-order Chebychev 
polynomial was used for the background pdf in order to better describe the edge of the phase space, which 
had a fairly sharp cut-off on the right-hand side of the $\omega$~signal peak. 

Figure~\ref{fig:Qfac_mass_histo} shows two examples of invariant $3\pi$ mass distributions for all 
linearly-polarized events in the energy bin $E_{\gamma} \in [1.1,\,1.2]$~GeV (left) and $E_{\gamma} 
\in[1.5,\,1.6]$~GeV (right), summed over all angles and polarization states.
The figure demonstrates the quality of the applied background-subtraction procedure: the total-mass 
distribution (black line) was nicely separated into a Voigtian mass distribution for the signal (red 
line), obtained by weighting each event with~$Q$, and a smooth polynomial background (blue line), 
obtained by weighting each event with ($1-Q$). At threshold, the choice for the background pdf did 
not always sufficiently constrain the likelihood fits. This occasionally manifested itself as small 
dip-like structures in the background mass distribution. Such effects were taken into account in
determining the systematic uncertainties associated with this method (see Section~\ref{ssec:sys_errors}). 

After applying all selection cuts and the event-based signal-background separation method, a total 
of 98,910 $p\,\omega$~events were retained from the entire data set using the combination 
of linear-beam polarization and transverse-target polarization, over the full photon energy range of 
1100\,-\,2100~MeV. From the corresponding data set using circular-beam polarization, 122,679~events 
were retained over the full incident-photon energy range of 1200\,-\,2800~MeV.

\section{Data Analysis\label{Section:Observables}}
The total cross section, $\sigma$, for $\omega$~photoproduction using a transversely-polarized target 
can be expressed in terms of the unpolarized cross section, $\sigma_0$, and a number of polarization 
observables as:
\begin{equation}
  \begin{split}
    \sigma\, =\, \sigma_0\,[\,1
      & - ~\bar{\delta}_l\,{\rm \Sigma}\,{\rm cos}\,2\phi \,+\,\bar{\Lambda}_{\rm t}\,{\rm T}\,{\rm sin}\,\alpha \\[0.5ex]
      & - ~\bar{\delta}_l\,\bar{\Lambda}_{\rm t}\,{\rm H}\,{\rm cos}\,\alpha\,{\rm sin}\,2\phi \\[0.5ex]
      & - ~\bar{\delta}_l\,\bar{\Lambda}_{\rm t}\,{\rm P}\,{\rm sin}\,\alpha\,{\rm cos}\,2\phi\,]\,,
   \end{split}
  \label{equ:reaction_rate_omega}
\end{equation}
where $\bar{\delta}_l$ denotes the average degree of linear-beam polarization (which was observed to be 
the same for `$+$' and `$-$' target polarizations), $\bar{\Lambda}_{\rm t}$ denotes the average target 
polarization (which was also observed to be the same for `$\parallel$' and `$\perp$' beam polarizations) 
and the azimuthal angle $\phi$~($\alpha$) is defined as the angle between the photon beam (target) 
polarization and the $\hat{x}_{\rm \,event}$-axis in the event frame, as shown in 
Fig.~\ref{fig:axes_and_angles_def}. Mathematically, 
\begin{equation}
 \phi= \phi_{\rm \,lab}^{\,p} - \pi - \phi_0,\quad\,\alpha = \pi - \phi_{\rm \,lab}^{\,p} 
            + \phi_{\rm \,offset}\,,
 \label{equ:angles_phi_alpha}
\end{equation}
which is also evident from the figure. Here, $\phi_{\rm \,lab}^{\,p}$ denotes the lab azimuthal angle of 
the recoiling proton and $\phi_0$~($\phi_{\rm \,offset}$) refers to the orientation of the photon-beam 
(transversely-polarized target) polarization with respect to the $\hat{x}_{\rm \,lab}$-axis in the 
laboratory frame. The definition of the angles and the polarization observables is analogous to the 
corresponding definition for the photoproduction of a single-pseudoscalar meson. When the beam 
polarization was set to $\parallel$ (or~$\perp$), then $\phi_0 = 0$ (or $\pi/2$)~rad. Similarly, 
$\phi_{\rm \,offset} = 2.025$ (or ($2.025 - \pi$))~rad when the target polarization was set to 
`$+$'~(or~`$-$'). These values in radians correspond to $\phi_{\rm \,offset} = 116.1^\circ$ 
and $-63.9^\circ$, respectively, as discussed in Section~\ref{Section:ExperimentalSetup}.
  
\begin{figure}[!t]
  \begin{center}
    \includegraphics[width=0.45\textwidth]{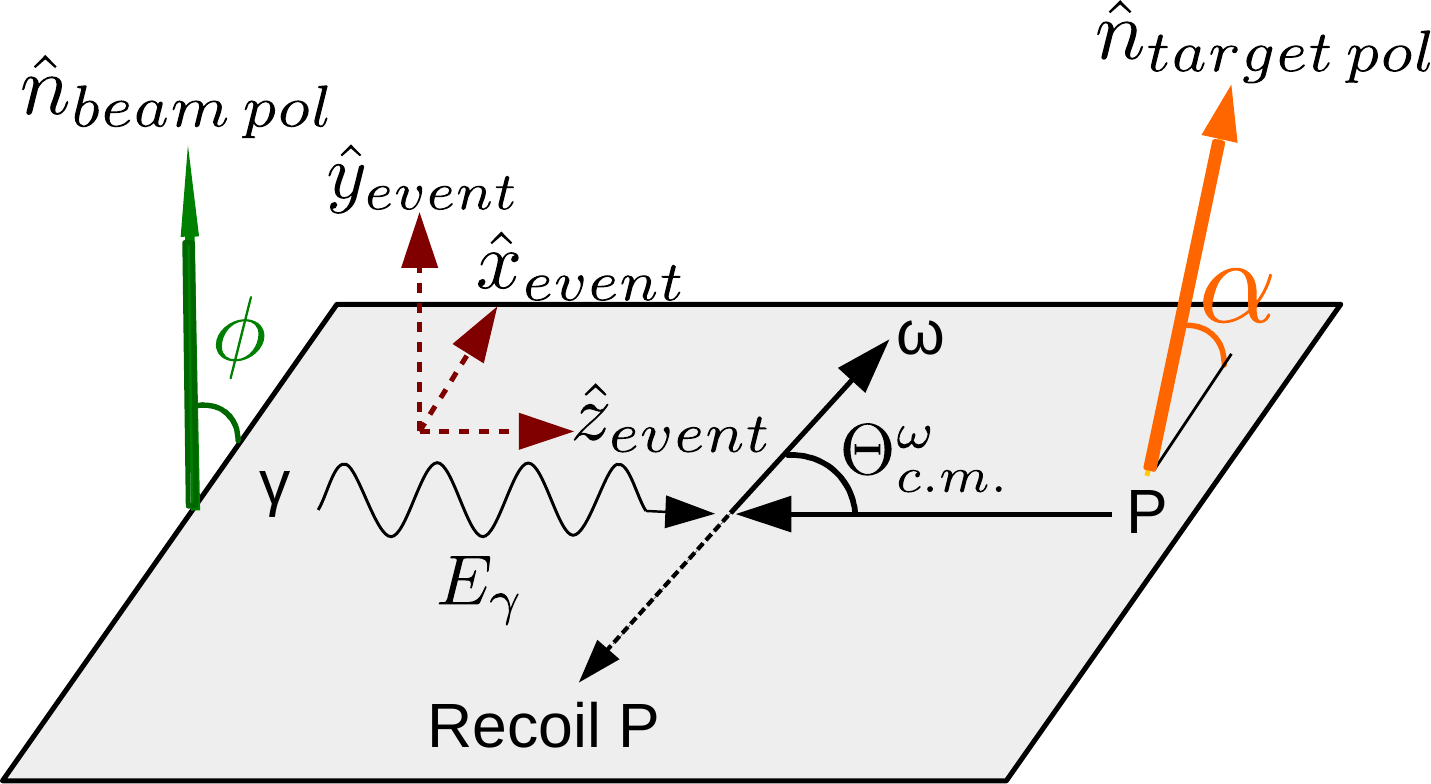}
    \caption{A diagram describing the kinematics of the reaction $\gamma p \to p\,\omega$. The
       plane represents the center-of-mass production plane defined by the initial photon and the recoil
       proton. The angle~$\Theta^{\,\omega}_{\rm \,c.m.}$ denotes the polar angle of the $\omega$~meson 
       in the event frame (or center-of-mass), defined in Section~\ref{Section:ExperimentalSetup}. Also 
       shown are the beam and target polarization orientations and the corresponding azimuthal angles, 
       $\phi$ and $\alpha$ (also in the center-of-mass frame).}
    \label{fig:KinematicPic_omega}
  \end{center}
\end{figure}

The total number of experimentally observed events is related to $\sigma$ as:
\begin{equation}
  N_{\rm \,data}\,=\,\Phi\,C\,\epsilon\,\sigma\,,
  \label{equ:Counts}
\end{equation}
where $\Phi$ is the incident photon flux, $C$ denotes the target cross-sectional area, and $\epsilon$ 
refers to the CLAS detector acceptance. The parameter~$\epsilon$ was observed to be independent of the 
relative orientation of the beam polarization with respect to the detector. Furthermore, the acceptance 
for the two target-polarization orientations was assumed to be very similar since the magnetic field of 
the holding magnet was fairly small. Small corrections of about 0.5~degrees or less were applied to the
azimuthal and polar angles of the detected final-state particles due to the effects of the holding field. 
More details on these corrections are available in Ref.~\cite{Roy:Dissertation}. 

For the extraction of asymmetries, the absolute value of the photon flux was not required. Rather, the 
ratios of fluxes between data sets with different beam-target polarizations were needed to effectively 
{\it unpolarize} the target in order to extract the beam asymmetry, $\Sigma$. The flux ratios were determined 
by using the information on the total number of reconstructed events from the polyethylene target, which 
was directly proportional to the photon flux. This target was chosen since the effects of the magnetic holding 
field were negligible at the target location. Events were also counted irrespective of topology so that the ratios 
were independent of the physics dynamics involved in the reaction specific to this analysis.
  
\subsection{Extraction of the photon-beam asymmetry, $\Sigma$\label{ssec:Likelihood}}  
Three independent kinematic variables were required to completely describe the event kinematics in 
$\omega$~photoproduction, as shown in Fig.~\ref{fig:KinematicPic_omega}. The following variables were chosen: 
the photon energy ($E_\gamma$), the polar angle of the $\omega$~meson in the center-of-mass frame
($\Theta_{\rm \,c.m.}^{\,\omega}$), and the azimuthal angle of the recoil proton ($\phi_{\rm \,lab}^{\,p}$)
in the laboratory frame (not shown in the figure). The observed modulation in the 
$\phi_{\rm \,lab}^{\,p}$~distribution was then used to extract the beam asymmetry at various ($E_\gamma$, 
$\Theta_{\rm \,c.m.}^{\,\omega}$) bins. An event-based maximum-likelihood fitting technique was implemented
to fit the angular modulations and extract $\Sigma$. This technique is considered more powerful than the 
conventional binning technique when the data suffer from low statistics since it uses information from every 
event, thereby preventing any loss of information due to binning. The method was based on the principles 
outlined in Ref.~\cite{Paterson:2016vmc}, which showed its application in a previous CLAS measurement. In 
this analysis, the likelihood (or the joint probability density) of obtaining the experimentally observed 
$\phi_{\rm \,lab}^{\,p}$~angular distribution was expressed in terms of $\Sigma$ as the only fit parameter 
(see Eq.~\ref{equ:A_sigma_1}-\ref{equ:weights}). To extract $\Sigma$ from the FROST data, the target 
polarization had to be removed (as detailed below). Maximizing the likelihood function then gave the 
most likely value for~$\Sigma$.
 
To nullify the effect of the target polarization to measure $\Sigma$, event samples with opposite target 
polarization but the same beam polarization were combined using appropriate scale (or normalization) 
factors. The number of $\parallel$~events, $N_\parallel$, after combining data sets with $\parallel$~beam 
polarization and different target polarizations (`$+$' or `$-$'), was given by:
\begin{equation}
N_{\parallel} \,=\, N_{\parallel}^+\,+\, N_1\, N_{\parallel}^-\,,
\label{equ:N_para_1}
\end{equation}
where $N_1$ was a normalization factor that depended on the photon flux, $\Phi_{\parallel}^+$ and 
$\Phi_{\parallel}^-$, and the average degrees of target polarization, $\bar{\Lambda}_{\rm \,t}^+$ and 
$\bar{\Lambda}_{\rm \,t}^-$, of the two data sets:
\begin{equation}
N_1 \,=\, \frac{\Phi_{\parallel}^+\,\bar{\Lambda}_{\rm \,t}^+}
               {\Phi_{\parallel}^-\,\bar{\Lambda}_{\rm \,t}^-}\,.
\label{equ:norm1}
\end{equation}
Substituting Eq.~\ref{equ:reaction_rate_omega} and~\ref{equ:Counts} into Eq.~\ref{equ:N_para_1} gave:
\begin{equation}
\begin{split}
N_{\parallel} \,=\, & \,\Phi_\parallel^+ \, C \,\epsilon\, \sigma_0\, (1 + \bar{\Lambda}_R)\,\{ 1 
    - \bar{\delta}_\parallel \,\Sigma \,{\rm cos} \,2\phi^{\,p}_{\rm \,lab} \}\\[0.5ex]
    \,=\, & \,\Phi_\parallel^+\,\epsilon\,\sigma_\parallel\,,
\end{split}
\label{equ:N_para_2}
\end{equation}
where $\bar{\Lambda}_R$ was defined as 
$\bar{\Lambda}_R \,=\,\bar{\Lambda}_{\rm \,t}^+ \,/\,\bar{\Lambda}_{\rm \,t}^-$.

Similarly, the number of $\perp$ events, $N_{\perp}$, after combining data sets with $\perp$~beam 
polarization and different target polarizations was given by:
\begin{equation}
\begin{split}
N_{\perp} \,=\, & \,\Phi_\perp^+\, C \,\epsilon\,\sigma_0\, (1 + \bar{\Lambda}_R)\, \{ 1 
    + \bar{\delta}_\perp \, \Sigma \,{\rm cos} \,2\phi^{\,p}_{\rm \,lab} \}\\[0.5ex]
    \,=\, & \,\Phi_\perp^+\,\epsilon\,\sigma_\perp\,.
\end{split}
\label{equ:N_perp_2}
\end{equation}

The asymmetry between $\parallel$~and $\perp$~data could then be expressed as:
\begin{equation}
\begin{split}
 A & \,=\, \frac{N_{\parallel}\,-\,N_{\perp}\,}{N_{\parallel}\,+\,N_{\perp}}
        \,=\, \frac{A^\prime \,+\, \Delta \Phi}{1\,+\,A^\prime\,\Delta \Phi}~,
\end{split}
\label{equ:A_sigma_1}
\end{equation}
where
\begin{equation}
\begin{split}
A^\prime &\,=\, \biggl(\frac{\sigma_{\parallel}\,-\,\sigma_{\perp}} 
     {\sigma_{\parallel}\,+\,\sigma_{\perp}}\biggr)
   \,=\, \frac{-\,\tilde{\delta}_l\,\Sigma\,{\rm cos}\,2\phi^{\rm \,p}_{\rm \,lab}}
   {1\,-\,\tilde{\delta}_l\,\Delta\delta_l\,\Sigma\,{\rm cos}\, 2\phi^{\rm \,p}_{\rm \,lab}}~,\\[1.5ex]
   \Delta \Phi &\,=\, \frac{\Phi_{\parallel}^+\,-\,\Phi_{\perp}^+} {\Phi_{\parallel}^+\,+\,\Phi_{\perp}^+}
   \quad {\rm and}\\[1.5ex]
   \tilde{\delta_l} &\,=\, \frac{ \bar{\delta}_{\parallel}\,+\,\bar{\delta}_{\perp}}{2}\,,\qquad
   \Delta\delta_l \,=\, \frac{\bar{\delta}_{\parallel}\,-\,\bar{\delta}_{\perp}} 
   {\bar{\delta}_{\parallel}\,+\,\bar{\delta}_{\perp}}~.
\end{split}
\label{equ:A_sigma_2}
\end{equation}

The likelihood of obtaining the observed angular distribution in $\phi^{\,p}_{\rm \,lab}$ in any 
kinematic bin, using $A$ from Eq.~\ref{equ:A_sigma_1}-\ref{equ:A_sigma_2}, was given by:
\begin{equation}
\begin{split}
-\text{ln}\,L \,=\, -\sum_{i=1}^{N_{\rm{\,total}}}\,w_i & \,\text{ln}\,(P\,(\rm{event_{\,i}})\,)\,,\\[0.5ex]
\end{split}
\end{equation}
\begin{equation*}
\begin{split}
 & {\rm{where}}~P\,(\text{event}_{\,i})=
\begin{cases}
 ~\frac{1}{2}\,(1\,+\,A) & \text{for $\parallel$ events}\,,\\[1ex]
 ~\frac{1}{2}\,(1\,-\,A) & \text{for $\perp$ events}\,,
\end{cases}
\end{split}
\label{equ:lnL_Sigma}
\end{equation*} 
and $N_{\rm \,total}$ denotes the sum of events over all four beam-target polarization settings used in 
that kinematic bin. The weight for each event depended on its $Q_{\rm \,event}$ and the normalization 
factor for the corresponding data set. From the above discussion, the weight of the $i^{th}$ event was 
given by:
\begin{equation}
w_i \,=\,
\begin{cases}            
  ~Q_i, & \text{for}~(\parallel,\,+)~{\textrm{or}}~(\perp,\,+)~{\textrm{events}}\,,\\[1.5ex]
  ~Q_i\,\frac{\Phi^{+}_{\parallel}\,\bar{\Lambda}^+}{\Phi^{-}_{\parallel}\,\bar{\Lambda}^-}, 
           & \text{for}~(\parallel,\,-)~{\textrm{events}}\,,\\[2.0ex]
  ~Q_i\,\frac{\Phi^{+}_{\perp}\,\bar{\Lambda}^+}{\Phi^{-}_{\perp}\,\bar{\Lambda}^-}, 
           & \text{for}~(\perp,\,-)~{\textrm{events}}\,.
\end{cases}
\label{equ:weights}
\end{equation}
Minimizing $-$ln\,$L$ yielded the value and the statistical uncertainty of the polarization observable 
$\Sigma$. This was performed at every ($E_\gamma$, $\Theta_{\rm \,c.m.}^{\,\omega}$)~bin. The MINUIT 
software package~\cite{cern:minuit} was used for the minimization.

\subsection{Extraction of the target asymmetry, $T$\label{ssec:TargetAsymmetry}}  
The target asymmetry $T$ was extracted from data using a transversely-polarized target and an incident 
circularly-polarized photon beam. The same likelihood technique described in subsection~\ref{ssec:Likelihood} 
was used to determine this polarization observable. Since the incident photons were polarized, this beam 
polarization had to be nullified. 

The number of events with target polarization `$+$', $N^+$, after combining events with different helicity 
states, was given by:
\begin{equation}
N^+ \,=\, N_\rightarrow^+\,+\, C_\leftarrow^+\, N_\leftarrow^+\,,
\label{equ:N_plus_1}
\end{equation}
where the normalization factor was $C^+_\leftarrow =1$ since the helicity state flipped rapidly and the 
events were not separated into different data sets. By substituting the value of $C^+_\leftarrow$ into 
Eq.~\ref{equ:N_plus_1} and using Eq.~\ref{equ:reaction_rate_omega} and \ref{equ:Counts}, the number $N^+$ 
was given by:
\begin{equation}
\begin{split}
N^+ \,=\, & \,2\,\Phi^+\,\epsilon\, \sigma_0\,\,\{ 1 + \bar{\Lambda}_{\rm \,t}^+
       \,T \,{\rm sin}\,(\pi - \phi^{\,p}_{\rm \,lab} + 2.025)\}\\[0.5ex]
    \,=\, & \,\Phi^+\,\epsilon\,\sigma^+\,,
\end{split}
\label{equ:N_plus_2}
\end{equation}
where $\Phi^+$ was the flux for the data set with target polarization `$+$'.

Similarly, the number of events with target polarization `$-$', $N^-$, after combining events with 
different helicity states, was given by:
\begin{equation}
\begin{split}
N^- \,=\, & \,2\,\Phi^-\,\epsilon\, \sigma_0\,\,\{ 1 - \bar{\Lambda}_{\rm \,t}^-
       \,T \,{\rm sin}\,(\pi - \phi^{\,p}_{\rm \,lab} + 2.025)\}\\[0.5ex]
    \,=\, & \,\Phi^-\,\epsilon\,\sigma^-\,,
\end{split}
\label{equ:N_minus_2}
\end{equation}
where $\Phi^-$ was the flux for the data set with target polarization `$-$'.

The asymmetry between target `$+$'~and `$-$'~data could then be expressed as:
\begin{equation}
 A \,=\, \frac{A^\prime \,+\, \Delta \Phi}{1\,+\,A^\prime\,\Delta \Phi}~,
\label{equ:A_T_1}
\end{equation}
where
\begin{equation}
\begin{split}
A^\prime &\,=\, \biggl(\frac{\sigma^+\,-\,\sigma^-} {\sigma^+\,+\,\sigma^-}\biggr)\\[1.5ex]
   & \,=\, \frac{\bar{\Lambda}_{\rm \,t}\,T\,{\rm sin}\,(\pi\,-\,\phi^{\rm \,p}_{\rm \,lab}\,+\,2.025)}
   {1\,+\,\bar{\Lambda}_{\rm \,t}\,\Delta\Lambda_{\rm \,t}\,T
   \,{\rm sin}\,(\pi\,-\,\phi^{\rm \,p}_{\rm \,lab}\,+\,2.025)}~,
\end{split}
\end{equation}
\begin{equation*}
\begin{split}
   \Delta \Phi &\,=\, \frac{\Phi^+\,-\,\Phi^-} {\Phi^+\,+\,\Phi^-}\quad {\rm and}\\[1.5ex]
   \bar{\Lambda}_{\rm \,t} &\,=\, \frac{\bar{\Lambda}_{\rm \,t}^+\,+\,\bar{\Lambda}_{\rm \,t}^-}{2}\,,\qquad
   \Delta\Lambda_{\rm \,t} \,=\, \frac{\bar{\Lambda}_{\rm \,t}^+\,-\,\bar{\Lambda}_{\rm \,t}^-}
   {\bar{\Lambda}_{\rm \,t}^+\,+\,\bar{\Lambda}_{\rm \,t}^-}~.
\end{split}
\label{equ:A_T_2}
\end{equation*}

The likelihood of obtaining the observed angular distribution in $\phi^{\,p}_{\rm \,lab}$ in any 
kinematic bin, using $A$ from Eq.~\ref{equ:A_T_1}, was given by:
\begin{equation}
\begin{split}
-\text{ln}\,L \,=\, -\sum_{i=1}^{N_{\rm{\,total}}}\,w_i & \,\text{ln}\,(P\,(\rm{event_{\,i}})\,)\,,\\[0.5ex]
\end{split}
\end{equation}
\begin{equation*}
\begin{split}
  & {\rm{where}}~P\,(\text{event}_{\,i})=
\begin{cases}
 ~\frac{1}{2}\,(1\,+\,A) & \text{for `$+$' events}\,,\\[1ex]
 ~\frac{1}{2}\,(1\,-\,A) & \text{for `$-$' events}\,,
\end{cases}
\end{split}
\label{equ:lnL_T}
\end{equation*} 
and $N_{\rm \,total}$ denotes the sum of events over the two target-polarization settings used in that  
kinematic bin. The weight of the $i^{th}$~event was $Q_i$ for all events. The observable $T$ was then 
extracted by minimizing -ln\,$L$.

\begin{figure*}[t]
  \begin{center}
    \includegraphics[width=1.0\textwidth,height=0.54\textheight]{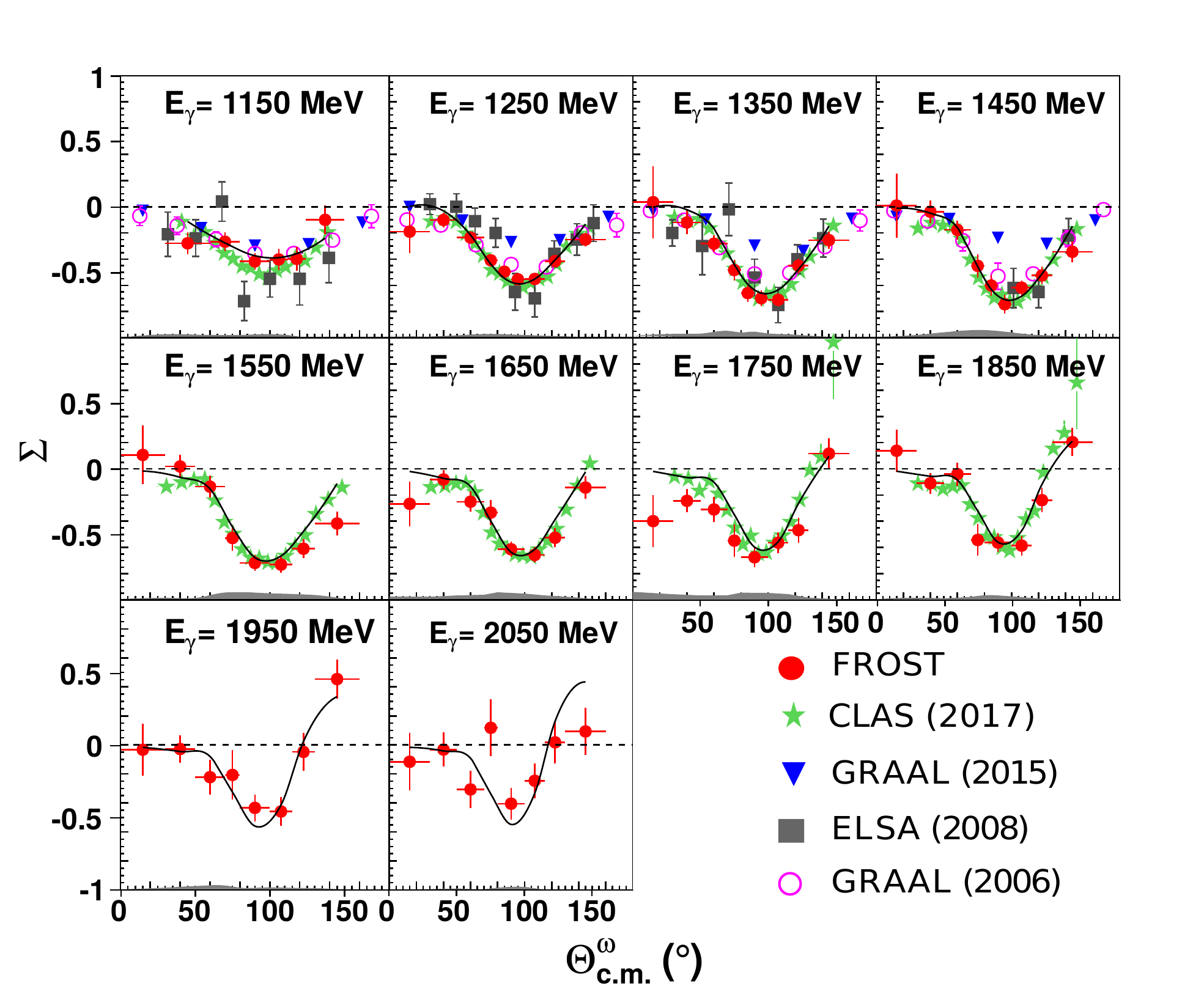}
    \caption{Results for the beam asymmetry, $\Sigma$, using a linearly-polarized photon 
       beam and an unpolarized target in the reaction $\gamma p\to p\,\omega$. The data are shown for 
       the energy range $E_{\gamma} \in [1.1,\,2.1]$~GeV in 100-MeV wide bins. The energy given for 
       each panel represents the energy of the bin center. The FROST results (red circles {\large\color{red} 
       $\bullet$}) are compared with previously published results from the GRAAL Collaboration in 2006 
       using the $\pi^+\pi^-\pi^0$~decay mode~\cite{Ajaka:2006bn} (magenta open circles 
       {\large\color{magenta} $\circ$}) and in 2015 using a weighted average of results from the 
       $\pi^+\pi^-\pi^0$ and $\pi^0\gamma$~decay modes in the energy range $E_{\gamma} \in [1.1,\,1.4]$~GeV 
       and from the radiative decay mode alone in the $[1.4,\,1.5]$~GeV $E_{\gamma}$~bin~\cite{Vegna:2013ccx}
       (blue inverted triangles {\color{blue} $\blacktriangledown$}), the CBELSA/TAPS Collaboration in 2008 
       using the radiative decay channel~\cite{Klein:2008aa} (gray squares {\color{gray05} $\blacksquare$}),
       and the CLAS Collaboration in 2017~\cite{Collins:2017vev} (green stars {\large\color{green} $\star$}). 
       The gray band at the bottom of each panel represents the absolute systematic uncertainties of our 
       results due to the background subtraction. The horizontal bars of the FROST data points indicate the 
       angular range they cover. The black solid line denotes the BnGa-PWA solution~\cite{Anisovich:2017}.}
    \label{fig:Sigma_all_E}
  \end{center}
\end{figure*}

\section{Results\label{Section:Results}}
This section presents the experimental results for the beam asymmetry, $\Sigma$, and the target asymmetry, 
$T$, in the photoproduction of a single $\omega$~meson off the proton. The $\Sigma$~observable can be 
compared with published results from various experiments and excellent agreement is observed, in particular 
with recent CLAS measurements using a liquid-hydrogen target. Since extracting single-spin observables from 
double-polarization data is challenging, this good agreement for $\Sigma$ provides confidence in the quality 
of the first-time measurements of the associated target asymmetries.

\begin{figure*}[t]
  \begin{center}
    \includegraphics[width=1.0\textwidth,height=0.49\textheight]{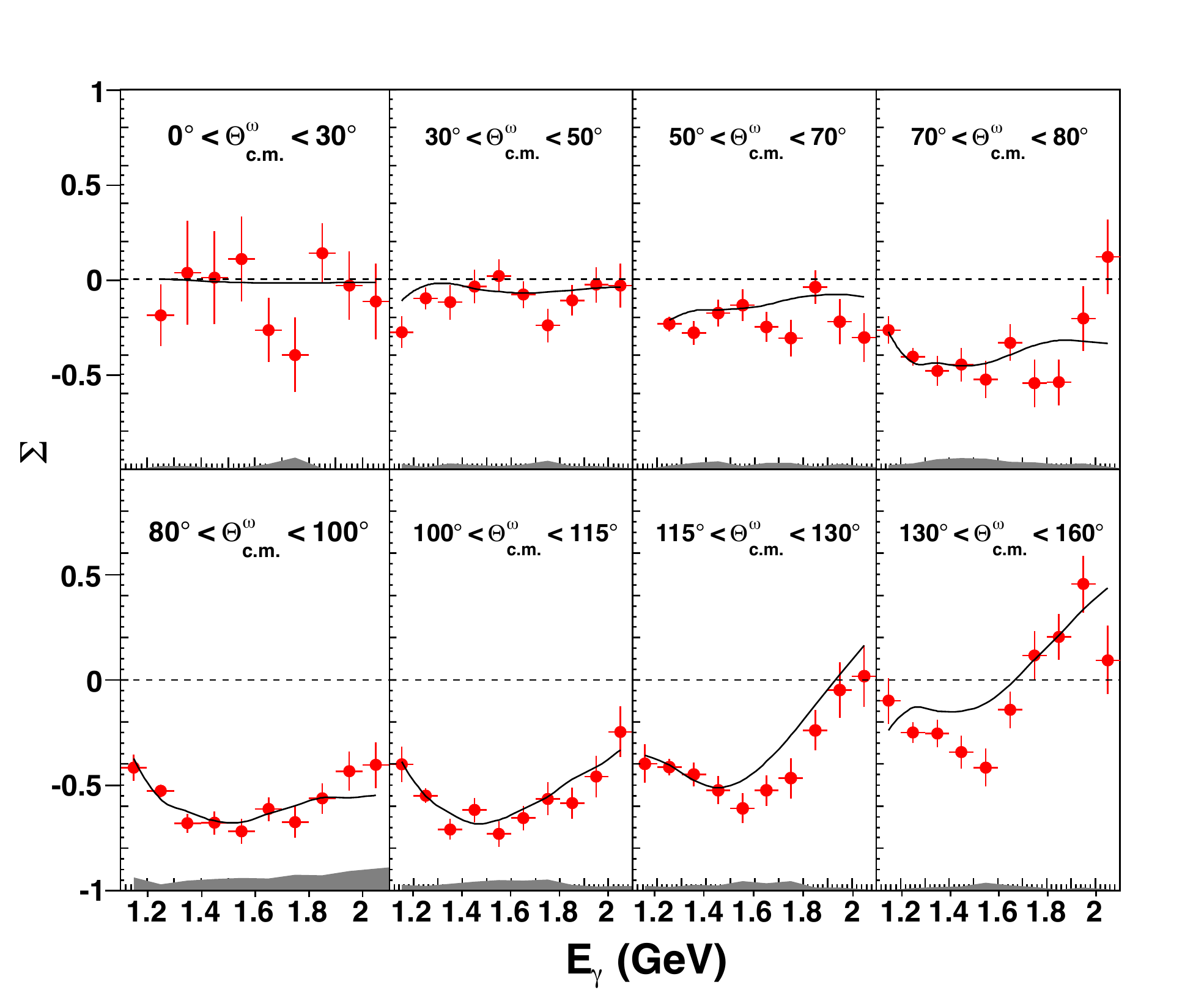}
    \caption{Results for the beam asymmetry, $\Sigma$, using a linearly-polarized photon beam and an 
       unpolarized target in the reaction $\gamma p\to p\,\omega$. The data are shown in bins of
       $\Theta_{\rm \,c.m.}^{\,\omega}$ versus incident-photon energy for the range of 
       $E_{\gamma}\in [1.1,\,2.1]$~GeV. The gray band at the bottom of each panel represents the absolute 
       systematic uncertainties of our results due to the background subtraction. The horizontal bars of the 
       FROST data points indicate the angular range they cover. The black solid line denotes the BnGa-PWA 
       solution~\cite{Anisovich:2017}.}
    \label{fig:Sigma_vs_E}
  \end{center}
\end{figure*}

\subsection{The Beam Asymmetry $\Sigma$}
Figure~\ref{fig:Sigma_all_E} shows the results for $\Sigma_\omega$ in the photoproduction reaction 
$\gamma p\to p\omega$ (Eq.~\ref{equ:omega_reaction}) including the statistical uncertainties for each data 
point from FROST (shown as red circles) as a function of $\Theta_{\rm \,c.m.}^{\,\omega}$. The data points 
are given for 10~energy bins in the incident-photon energy range $[1.1,\,2.1]$~GeV; Each energy bin is 
$0.1$-GeV wide. The numerical values for the data presented in Fig.~\ref{fig:Sigma_all_E} including the 
statistical and systematic uncertainties are available in Ref.~\cite{PRC:SupplementalMaterial}. The very 
forward and backward $\Theta_{\rm \,c.m.}^{\,\omega}$~angles had low statistics owing to poor CLAS acceptance. 
Therefore, a variable binning scheme for this angle range was chosen such that the bins at the very forward 
and backward regions are wider than the bins in the central region.

The FROST data points above $1.9$~GeV in incident photon energy represent first-time measurements. Also 
shown in the figure are published results from other experiments: two sets of results from the GRAAL 
Collaboration~\cite{Ajaka:2006bn} (2006 data - magenta open circles) and~\cite{Vegna:2013ccx} (2015 
data - blue inverted triangles). The GRAAL-2006 data cover the energy range from the reaction threshold up 
to $1.5$~GeV and were extracted from the $\omega\to\pi^+\pi^-\pi^0$ decay mode. The GRAAL-2015 data 
cover the same energy range but represent a statistics-weighted average of results obtained from the 
$\pi^+\pi^-\pi^0$ and the radiative $\pi^0\gamma$ decay modes, with the exception of the $1.45$~GeV 
photon energy bin where the results were obtained from the radiative decay mode only. The CBELSA/TAPS 
Collaboration published results from the $\omega\to\pi^0\gamma$ decay mode in 2008 for energies up 
to $1.7$~GeV~\cite{Klein:2008aa} (gray squares). Also shown are recent results from the CLAS 
Collaboration~\cite{Collins:2017vev} from a liquid-hydrogen experiment (green stars). 
These latter data from CLAS are in excellent agreement with the new data from this analysis and serve as a 
validation for the first-time measurements of the $\omega$~target asymmetry presented in the following 
section.

The overall agreement of the angular distributions from all experiments ranges from fair to good with some 
more serious discrepancies in certain $\Theta_{\rm \,c.m.}^{\,\omega}$~bins. For example, the CBELSA/TAPS 
data points tend to be bigger in magnitude than the GRAAL 2006 results, particularly for the center angles, 
$\Theta_{\rm \,c.m.}^{\,\omega} \in [80,\,120]^\circ$, of the first two energy bins. The GRAAL Collaboration 
aimed at resolving this issue with additional measurements but the results published in 2015 exhibited even 
greater inconsistencies with the previous measurements, especially between the two GRAAL measurements 
themselves. The more recent results appear to be significantly smaller in magnitude in the central region 
around $\Theta_{\rm \,c.m.}^{\,\omega} = 90^\circ$.

In the lower-energy range below 1.5~GeV, the CLAS results can be compared with the previously published 
data. They are in very good agreement with the GRAAL 2006 and in fair agreement with the GRAAL 2015 results 
close to the threshold. The CBELSA/TAPS data points suffer from significantly larger statistical uncertainties
but the agreement with the CLAS results is fair and mostly within uncertainties. All of this provides confidence 
in the new CLAS-FROST data and also resolves the inconsistency between the two GRAAL measurements in favor 
of the 2006 results.


Figure~\ref{fig:Sigma_vs_E} shows the beam asymmetry as a function of the incident-photon energy for 
different $\Theta_{\rm \,c.m.}^{\,\omega}$~bins. The first angle bin, $\Theta_{\,c.m.}^{\,\omega} \in 
[0,\,30]^\circ$, suffered from low statistics at all energies. However, the results in the subsequent 
angle bins clearly show that the overall shape of the beam asymmetry with respect to energy changes 
noticeably upon moving from forward to backward angles. The asymmetry is small and almost consistent 
with zero across the entire energy range for $[30,\,50]^{\circ}$, whereas it grows bigger in the 
successive angle bins, reaching a value of about 0.55 in the $[80,\,100]^{\circ}$~angle bin.

\subsection{The Target Asymmetry $T$}
\begin{figure*}[!ht]
  \begin{center}
    \includegraphics[width=1.0\textwidth]{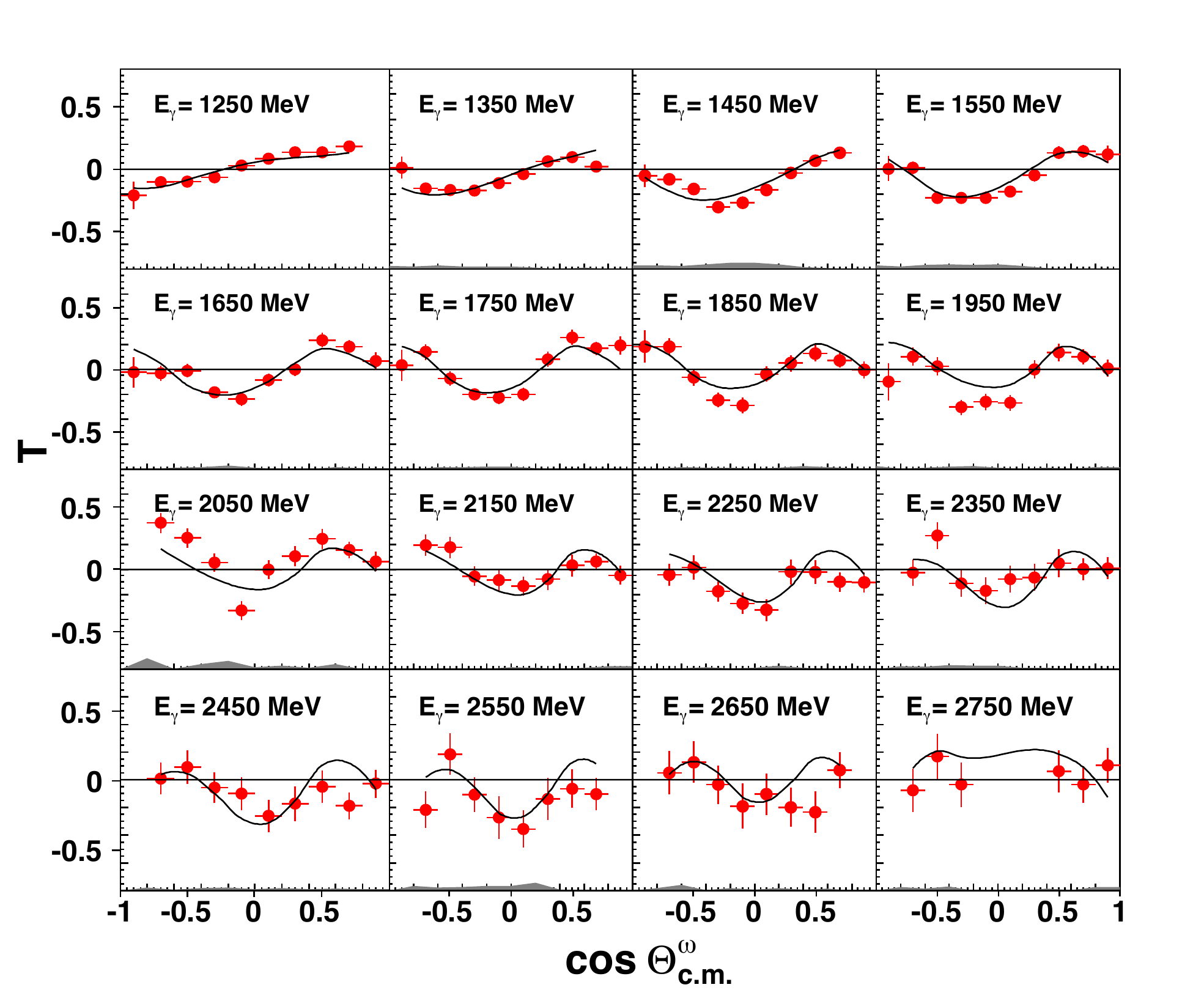}
    \caption{Results for the target asymmetry, $T$, using a transversely-polarized target 
       in the reaction $\gamma p\to p\,\omega$. The data are shown for the energy range $E_{\gamma} \in 
       [1.2,\,2.8]$~GeV in 100-MeV wide bins. The gray band at the bottom of each panel represents the 
       absolute systematic uncertainties of our results due to the background subtraction. The horizontal 
       bars of the FROST data points indicate the angular range they cover. The  black solid line denotes
       the BnGa-PWA solution~\cite{Anisovich:2017}.}
    \label{fig:T_all_E}
  \end{center}
\end{figure*}

Figure~\ref{fig:T_all_E} shows the results for the target asymmetry in the photoproduction reaction 
$\gamma p\to p\omega$ (Eq.~\ref{equ:omega_reaction}) including the statistical uncertainties for each 
data point from FROST as a function of cos\,$\Theta_{\rm \,c.m.}^{\,\omega}$. The data points are given 
for 16~energy bins in the incident-photon energy range $[1200,\,2800]$~MeV; Each energy bin is $100$-MeV 
wide. The numerical values for the data presented in Fig.~\ref{fig:T_all_E} including the statistical 
and systematic uncertainties are available in Ref.~\cite{PRC:SupplementalMaterial}. The observable exhibits 
rich structures and acquires large values of about 0.3\,-\,0.4 around cos\,$\Theta_{\rm \,c.m.}^{\,\omega} = 0$ 
over a large energy range.

\subsection{Systematic uncertainties\label{ssec:sys_errors}}
The individual contributions to the overall systematic uncertainty for each observable that were studied 
in this analysis are listed in Table~\ref{tab:Sys_err}. The absolute systematic uncertainty due to the background 
subtraction is shown as an error band at the bottom of each distribution in 
Figs.~\ref{fig:Sigma_all_E}\,-\,\ref{fig:T_all_E}. 
The fractional uncertainties were added in quadrature and the totals are given in Table~\ref{tab:Sys_err}.

A major contribution came from the event-based background-subtraction technique. 
To estimate this contribution to the overall systematic uncertainty, the $Q$~value of each event was 
increased by $\sigma_Q$ and the beam asymmetry was re-extracted. Here, $\sigma_Q$ denotes the fit 
uncertainty in the $Q$~value of the $i^{th}$ event. The change in the observable in each kinematic bin 
provided an absolute uncertainty in the observable due to this method. For the beam asymmetry, it was 
observed to be $8\,\%$ on average above $1300$~MeV in the incident photon energy. This procedure was 
based on the assumption that the chosen signal and background pdfs properly described the data. However, 
as mentioned in Section~\ref{ssec:Qfac}, the description was not always satisfactorily close to the 
$\omega$~photoproduction threshold. In such situations, a {\it dip-like} structure in the background 
distribution under the $\omega$~peak was observed. To estimate the systematics associated with this effect,
the background distribution was fitted with a second-order polynomial in the range 
$\omega_{\rm \,peak}\,\pm\,5\sigma$, where $\sigma$ was the width of the peak. The fractional difference 
between the original background and the fit in the range $\omega_{\rm \,peak}\,\pm\,2\sigma$ was determined 
to be about 5\,-\,7\,\% on average. To quantify the effect of this fractional difference on the final 
observables, the following strategy was employed: Since the background was under-estimated in the region 
$\omega_{\rm \,peak}\,\pm\,2\sigma$, equivalent to the signal being over-estimated, the $Q$~values of the 
events belonging to this mass range were changed by 
$\sigma_Q - 0.07\,Q$. The observable was then re-determined and the fractional difference between the 
original observable and the modified observable was quoted as the systematic uncertainty. It was determined 
to be $4.5\,\%$ on average in the energy range $E_{\gamma} \in [1.1,\,1.3]$~GeV.

\begin{table}[t!]
\caption{\label{tab:Sys_err} List of systematic uncertainties.}
\begin{center}
\begin{tabular}{lcc}
Source & & Systematic Uncertainty\\
\hline\hline
Background subtraction & ~~~ & given as gray band for each\\
 &  & distribution in Figs~\ref{fig:Sigma_all_E},\,\ref{fig:Sigma_vs_E},\,\ref{fig:T_all_E}\\
Beam-polarization & & $5\,\%$\\
Target-polarization & & $2\,\%$\\
Target-offset angle & & $2\,\%$\\
Normalization & &\\
\qquad beam asymmetry & & $5\,\%$\\
\qquad target asymmetry & & $2\,\%$\\
\hline\hline
Beam asymmetry & &\\
$\sigma_{\rm\,total}$ (fractional only) & & $\sim 7.5\,\%$\\\hline
Target asymmetry & &\\
$\sigma_{\rm\,total}$ (fractional only) & & $\sim 3.5\,\%$\\
\end{tabular}
\end{center}
\end{table}

The systematic uncertainty in the linear-beam polarization was evaluated to be $\sim 5\,\%$, a value which 
was also used in other CLAS analyses~\cite{Paterson:2016vmc,Dugger:2013crn}. The systematic uncertainty 
associated with the target polarization was determined to be $\sim 2\,\%$~\cite{Keith:2012ad}. 
To estimate the systematic uncertainty in the observable due to the target-offset angle, this angle was 
varied by its uncertainty of $\pm 0.4^\circ$
and the change in the re-extracted observable was examined. It was found to be $2\,\%$ on average.
 
For the measurement of the beam asymmetry, three factors were required to normalize the four 
linearly-polarized data sets, as can be seen from Eqs.~\ref{equ:norm1}-\ref{equ:A_sigma_2} 
(Section~\ref{ssec:Likelihood}):
\begin{equation}
N_1\,=\,\frac{\Phi^{+}_{\parallel}}{\Phi^{-}_{\parallel}}\,\bar{\Lambda}_R,\quad
N_2\,=\,\frac{\Phi^{+}_{\perp}}{\Phi^{-}_{\perp}}\,\bar{\Lambda}_R,\quad
N_3\,=\,\frac{\Phi^{+}_{\parallel}}{\Phi^{+}_{\perp}}~.
\end{equation}
The first two normalization factors were needed to {\it unpolarize} the target in the `$\parallel$' and 
`$\perp$' data sets, respectively. The third normalization factor was then required to normalize the 
corresponding `$\parallel$' and `$\perp$'  data sets (after the target was rendered {\it unpolarized}). 
The uncertainties in the normalization factors depended on the uncertainties in the flux ratios, which were 
obtained from the ratios of the numbers of reconstructed events originating from the polyethylene target. 
One way to estimate the systematic uncertainty in these ratios was to compare them with the ratios obtained 
from the carbon target. The results were found to differ by $2\,\%$ or less at all energies. Another way 
to check the systematics of this method was to use the direct information on the photon flux from the 
photon tagging system. Although this information was not available for the FROST data used in this analysis, 
it was available for FROST-g9a data, which utilized a circularly-polarized beam and a longitudinally-polarized 
target. The results differed again by only $\sim 2\,\%$ from those determined for the polyethylene target. 
The applied uncertainties of $2\,\%$ in the flux ratios as well as the uncertainty in the target 
polarization were used to evaluate the overall uncertainties in the normalization factors using standard 
error propagation. Since each normalization factor could be varied by $\pm\sigma$, all permutations were 
performed and the observable re-extracted. The change in the beam asymmetry was observed to be $5\,\%$ on 
average across all energies.

For the measurement of the target asymmetry using circularly-polarized data, only one factor was required 
to normalize data sets with opposite target polarization (Section~\ref{ssec:TargetAsymmetry}) and thus, 
the systematic uncertainty in the overall normalization was smaller than for the linearly-polarized data. 
Following the same procedure as for the beam asymmetry, the normalization factor was changed by 2\,\% and 
the observable re-extracted. An effect of $< 2\,\%$ was observed in the target asymmetry due to the 
normalization.

\section{Partial Wave Analysis\label{Section:PWA}}
The data presented here were included in a partial-wave analysis within the Bonn-Gatchina (BnGa) PWA 
framework. The scattering amplitudes in the BnGa analysis for the production and the decay of baryon 
resonances are constructed in the framework of the spin-momentum operator expansion method. 
The details of this approach are discussed in Ref.~\cite{Anisovich:2006bc}. The approach is relativistically 
invariant and allows for the combined analyses of different reactions imposing analyticity and unitarity directly. 
The BnGa~database takes into account almost all important data sets of photo- and pion-induced reactions, 
including three-body final states~\cite{Anisovich:2011fc}. A full description of the experimental 
database~\cite{BnGa:Database} goes beyond the scope of this paper. 

The BnGa group has recently reported on a PWA~\cite{Denisenko:2016ugz} of $\omega$~photoproduction 
data that was based on results from the CBELSA/TAPS Collaboration alone. The data sets and the relevant 
observables (d$\sigma$/d$\Omega$, SDMEs, $\Sigma$, $E$, and $G$), which were used in the PWA, are discussed 
in Refs.~\cite{Denisenko:2016ugz, Wilson:2015uoa}. The new BnGa-PWA solution, which includes data from the 
CLAS Collaboration, is shown in Figs.~\ref{fig:Sigma_all_E}\,-\,\ref{fig:T_all_E} as a solid line. The CLAS 
data include the polarization observables $\Sigma$ and $T$ (presented here), $F$, $P$, $H$, and $E$. More 
details on the PWA framework and branching ratios for $N^\ast$~decays into $N\omega$ will be discussed in 
a subsequent publication~\cite{Anisovich:2017}.

In the FROST $\gamma p\to p\omega$ data presented here, large beam asymmetries, as well as smaller but 
significantly non-zero target asymmetries are observed, which indicate significant $s$-channel contributions, 
in agreement with the expectation from the BnGa~PWA. Close to the reaction threshold, the leading partial waves 
are the $3/2^+$ and $5/2^+$~waves, which are identified with the $N(1720)\,3/2^+$ and the sub-threshold 
$N(1680)\,5/2^+$~nucleon resonances. Recent calculations that used an effective chiral Lagrangian 
approach~\cite{Zhao:2000tb} also found these two resonances to play a major role in $\omega$~photoproduction. 
In particular, the $N(1720)\,3/2^+$ was analyzed in the beam polarization asymmetries. The $3/2^+$~partial wave
is complex and multiple $3/2^+$~nucleon resonances likely contribute to our data around $W = 1.7 - 2.1$~GeV. 
The importance of the $3/2^+$~wave was also discussed in an earlier event-based PWA based on CLAS 
$\omega$~cross section data and unpolarized spin-density matrix elements alone~\cite{Williams:2009aa}. 
The BnGa~PWA finds indications for at least one more $3/2^+$~resonance around $W = 1.9$~GeV.

Toward higher energies, the $t$-channel contributions increase in strength and in the case of $\Sigma$, the 
linear-beam polarization allows for the separation of natural- from unnatural-parity exchange processes. The 
BnGa group has found that pomeron-exchange dominates over the smaller $\pi$-exchange across the presented 
energy range. Further $N^\ast$-resonance contributions are required to describe the data at and above 
center-of-mass energies of $W = 2$~GeV. The $1/2^-$, $3/2^-$, and $5/2^+$ partial waves play a significant 
role in the PWA solution. In addition to the $N(1680)\,5/2^+$ close to the threshold, a further structure around 
$W = 2$~GeV is observed, which is identified with the $N(2000)\,5/2^+$~state. The latter is listed as a one-star 
state in the RPP~\cite{Olive:2016xmw} and considered a missing baryon resonance.

A full discussion of the contributing resonances can be found 
in a forthcoming paper on the details of the PWA~\cite{Anisovich:2017}.

\section{Summary}
The photon-beam asymmetry~$\Sigma$ for the photoproduction reaction $\gamma p\to p\,\omega$ has been 
measured at Jefferson Laboratory using the CLAS spectrometer and the frozen-spin FROST target, covering 
the energy range from 1.1~to 2.1~GeV. The $\omega$~meson has been studied via its 
$\omega\to\pi^+\pi^-\pi^0$~decay. The high-quality FROST results are in overall fair agreement with 
previously published data (including CLAS) and help shed some light on earlier-observed discrepancies 
among the known data sets. Moreover, first-time measurements of the target asymmetry $T$ have been 
presented covering a large incident-photon energy range from 1.2~to 2.8~GeV. These data are rich in structures. 
The angular distributions change from an almost linear behavior close to the reaction threshold to a more 
oscillatory behavior at higher energies. The asymmetries acquire significant values of up to 
0.4, mostly around cos\,$\Theta_{\rm \,c.m.}^{\,\omega} = 0$.

\begin{acknowledgments}
The authors thank the technical staff at Jefferson Lab and at all the participating institutions for their 
invaluable contributions to the success of the experiment. This material is based upon work supported by 
the U.S. Department of Energy, Office of Science, Office of Nuclear Physics, under Contract No. 
DE-AC05-06OR23177. The group at Florida State University acknowledges additional support from the U.S. 
Department of Energy, Office of Science, Office of Nuclear Physics, under Contract No. DE-FG02-92ER40735.
This work was also supported by the US National Science Foundation, the State Committee of Science of 
Republic of Armenia, the Chilean Comisi\'{o}n Nacional de Investigaci\'{o}n Cientifica y Tecnol\'{o}gica 
(CONICYT), the Italian Istituto Nazionale di Fisica Nucleare, the French Centre National de la Recherche 
Scientifique, the French Commissariat a l'Energie Atomique, the Scottish Universities Physics Alliance 
(SUPA), the United Kingdom's Science and Technology Facilities Council, and the National Research Foundation 
of Korea, the Deutsche Forschungsgemeinschaft (SFB/TR110), and the Russian Science Foundation under grant 
16-12-10267.
\end{acknowledgments}

\end{document}